\documentclass[prc,aps,manuscript]{revtex4}

\usepackage{amsfonts}
\usepackage{amsmath}
\usepackage{graphicx}
\usepackage{dcolumn}
\usepackage{bm}

\begin{document}

\title{Stochastic features of dissipative large-amplitude dynamics and
nuclear fission}
\author{ \underline{V. M. \surname{Kolomietz}} and
S. V. \surname{Radionov} \thanks{%
Electronic address: sergey.radionov18@gmail.com}}
\affiliation{\textit{Institute for Nuclear Research, 03680 Kiev, Ukraine} }
\date{\today}

\begin{abstract}
Within a density matrix approach for nuclear many--body system,
it is derived non--Markovian Langevin equations of motion for
nuclear collective parameters, where memory effects are defined
by memory time. The developed stochastic approach is applied to
study both the nuclear descent from fission barrier to a scission
point and thermal diffusive overcoming of the barrier. The present
paper is partly a review of our results obtained earlier and contains
new results on the non--Markovian generalization of Kramers'
theory of escape rate and on time features of the collective
dynamics in the presence of periodic external modulation.
\end{abstract}

\maketitle

\section{Introduction}

Nuclear large-scale dynamics (nuclear fission, heavy ion collisions etc.) is
a good probe for the investigation of the complex time evolution of finite
Fermi systems. The principal question here is how macroscopic (collective)
modes of motion \cite{sije87,hamy88,hofm97} are affected by complex
microscopic (intrinsic) excitations of the many--body Fermi systems.
Due to dissipative and fluctuating character of the collective modes of motion,
manifesting in non--zero widths of the nuclear giant multipole resonances and
non--zero variances of the kinetic energy of the nuclear fission fragments,
one uses transport approaches of the Fokker--Planck and Langevin types
\cite{bosu93,bosu94,abay96,ayno82,kolo95,kora06,koab08,kora09} to the study
of nuclear large--amplitude dynamics. In general, basic equations for the
macroscopic collective variables are non--Markovian \cite{kora01,kosh04},
implying complex energy flow between the slow collective and fast intrinsic
degrees of freedom of the nuclear many--body system.

In the present paper, following the ideology of the random matrix theory
\cite{weiden80,bulgac01} we discuss how non--Markovian (memory) and
stochastic aspects of the nuclear fission dynamics are defined by
quantum--mechanical diffusion of energy in the space of occupancies
of complex many--body states. The general problem of
decay of a metastable state (like a decay of nuclear compound state)
has gotten a lot of attention in the functional integral approach
\cite{grab84,hanggi85} and in its application to the nuclear case
\cite{rumm02}. Quantum decay rate of the nuclear compound state has
been derived within the local harmonic approximation \cite{hofm83},
where also non--Markovian effects were considered. In spite of such
wide literature on the subject, the problem of classical activated
(escape) rate over the nuclear fission barrier in the presence of
the non--Markovian effects has been left without appropriate attention.
We are going to renew this deficiency by measuring a general
quantitative impact of memory effects on the classical thermal
rate and time characteristics of the nuclear fission dynamics.

The plan of the paper is as follows. In Sect.~II, we consider a many-body
dynamics by use the Zwanzig's projection technique and derive the basic
non--Markovian Langevin equations of motion within the cranking approach to
nuclear many--body dynamics. In Sect.~III, we discuss the application of
non--Markovian dynamics to the nuclear descent from the fission barrier, the
memory effect on the Kramers' diffusion over the fission barrier, the
stochastic penetration over oscillating barrier and the diffusion of
occupation probabilities within Landau-Zener approach. The non--Markovian
dynamics of nuclear Fermi liquid is considered in Sect.~IV. Summary and
conclusions are given in Sect.~V.

\section{Many-body dynamics in a moving frame}

We assume that dynamics of nuclear many--body system can be described as a
coupled motion of several slow macroscopic (collective) modes and intrinsic
nucleonic ones. Slow collective modes are treated in terms of a set of
classical time--dependent variables $q(t)\equiv
\{q_{1}(t),q_{2}(t),...,q_{N}(t)\}$. The fast intrinsic modes are described
quantum mechanically through the Liouville equation for the density matrix
operator $\hat{\rho}$,
\begin{equation}
\frac{\partial \hat{\rho}(t)}{\partial t}+i\hat{L}(t)\hat{\rho}(t)=0,
\label{L}
\end{equation}%
where $\hat{L}$ is the Liouville operator defined as
\begin{equation}
\hat{L}\hat{\rho}=\frac{1}{\hbar }\left[ \hat{H},\hat{\rho}\right] .
\label{H}
\end{equation}%
Here $\hat{H}(q)$ is the nuclear many--body Hamiltonian. We introduce a
moving (adiabatic) basis of the Hamiltonian $\hat{H}(q)$,
\begin{equation}
\hat{H}(q)\Psi _{n}(q)=E_{n}(q)\Psi _{n}(q),  \label{adiabbas}
\end{equation}%
determined by a set of eigenfunctions $\Psi _{n}(q)$ and eigenenergies $%
E_{n}(q)$ for each fixed value of the macroscopic variables $q(t)$. Within
this basis one can introduce a non--diagonal, $\rho _{nm}$, and diagonal, $%
\rho _{nn}$, parts of the density matrix through the relations,
\begin{equation}
\rho _{nn}=\langle \Psi _{n}|\hat{\rho}|\Psi _{n}\rangle ,~~~~~\rho
_{nm}=\langle \Psi _{n}|\hat{\rho}|\Psi _{m}\rangle ,  \label{rhorho}
\end{equation}%
and whose time evolution may be determined within the Zwanzig's projection
technique \cite{zwan60}. The final results read, see Refs. \cite%
{kolo95,kora10},
\begin{eqnarray}
\rho _{nm}(t)=\rho_{nm}(t=0)- i\sum_{j=1}^{N}\int_{0}^{t}dt^{\prime }\dot{q}%
_{j}(t^{\prime })\frac{\mathrm{exp}[-i\omega _{nm}(t-t^{\prime })]}{\omega
_{nm}}  \notag \\
\times \left[ h_{j,mn}(t^{\prime })\rho _{nn}(t^{\prime
})-h_{j,nm}(t^{\prime })\rho _{mm}(t^{\prime })\right] ,  \label{rnm}
\end{eqnarray}%
and its diagonal part,
\begin{eqnarray}
\frac{\partial \rho _{nn}(t)}{dt} &=&\frac{2}{\hbar ^{2}}\sum_{i,j=1}^{N}%
\dot{q}_{i}(t)\int_{0}^{t}dt^{\prime }\dot{q}_{j}(t^{\prime })\sum_{m\neq
n}h_{i,nm}(t)h_{j,mn}(t^{\prime })\frac{\mathrm{cos}[\omega
_{nm}(t-t^{\prime })]}{\omega _{nm}^{2}}  \notag \\
&&\times \left[ \rho _{mm}(t^{\prime })-\rho _{nn}(t^{\prime })\right] .
\label{rnn}
\end{eqnarray}%
Here, $\omega _{nm}=(E_{n}-E_{m})/\hbar $ and matrix elements $%
h_{i,nm}=\left( \partial \hat{H}/\partial q_{i}\right) _{nm}$ measure the
coupling between the quantum nucleonic and the macroscopic collective
subsystems. Note that the equations (\ref{rnm}) and (\ref{rnn}) were derived
in the weak--coupling limit \cite{kora10}. This condition can be violated
near the avoided crossings of two nearest levels. One can go beyond the
second--order perturbation result (\ref{rnm})--(\ref{rnn}) by considering
the Landau--Zener model of two crossing levels as shall be done below.

A complexity of a quantum system is understood as the absence of any special
symmetries in a system. Such a system is expected to have some universal
statistical properties which can be modeled by the random matrix ensembles.
Within the random matrix approach~\cite{goel65}, we will average the
right--hand side of master equation~(\ref{rnn}) over suitably chosen
statistics of the randomly distributed matrix elements $h_{i,nm}$ and the
energy spacings $E_{n}-E_{m}$. First, we perform the ensemble averaging over
the matrix elements. They are treated as complex random numbers with the
real and the imaginary parts which are independently Gaussian distributed,
and with \cite{wilk89}
\begin{equation}
\overline{h_{i,nm}(q)h_{j,n^{\prime }m^{\prime }}^{\ast }(q^{\prime })}%
=\delta _{nn^{\prime }}\delta _{mm^{\prime }}\sigma
_{ij}^{2}(E_{n},E_{m},q+q^{\prime })Y_{ij}(|q-q^{\prime }|),  \label{hh}
\end{equation}%
where $Y_{ij}(|q-q^{\prime }|)$ is a correlation function, measuring how
strong the ensemble averaged matrix elements correlate at different
collective deformations $q$ and $q^{\prime }$, and $\sigma
_{ij}^{2}(E_{n},E_{m},q+q^{\prime })$ is the energy distribution of the
squared matrix elements. It is rather clear that at high excitation energies
the coupling matrix elements between the complex many--body states should
drop out with increasing energy distance between them. In order to
characterize the energy distribution of the matrix elements, we introduce
the strength of the distribution $\sigma _{0,ij}^{2}$ and its width $\Gamma
_{ij}$ assuming
\begin{equation}
\sigma _{ij}^{2}(E_{n},E_{m},q+q^{\prime })=\frac{\sigma
_{0,ij}^{2}(E_{n}+E_{m},q+q^{\prime })}{\sqrt{\Omega(E_{n})\Omega(E_{m})}
\Gamma _{ij}}f_{ij}(|E_{n}-E_{m}|/\Gamma _{ij}),~
\end{equation}%
where $\Omega(E)$ is the average level--density at given excitation energy $%
E $ and $f(E)$ is a shape of the energy distribution. The correlations of
the matrix elements, existing at different values of the macroscopic
parameter, $q$ and $q^{\prime }$, are measured with the help of the
correlation function $Y_{ij}$. Since the energy correlations between two
different states $n$ and $m$ drop out with the rise of a distance between
them, it is rather obvious that $f_{ij}\rightarrow 0$ with $%
|E_{n}-E_{m}|/\Gamma_{ij}\rightarrow \infty $ and $f_{ij}(E)\sim 1$ at $%
|E_{n}-E_{m}|/\Gamma \ll 1$. As is discussed in Refs.~\cite{brne79,zelev95},
for nuclear shell--model Hamiltonians typical values of the spreading
width $\Gamma$ are of several (tens) MeV.

The energy spacings part of the ensemble averaging procedure is defined
through the two--level correlation function, $R(|E_{n}-E_{m}|\Omega(E))$,
that is the probability density to find the state $m$ with energy $E_{m}$
within the interval $[E_{m},E_{m}+dE_{m}]$ at the average distance $%
|E_{n}-E_{m}|$ from the given state $n$ with energy $E_{n}$. In the case of
quite dense spectrum of the adiabatic states $n$, it is more convenient to
use continuous energy variables $E$\ and $e$, that measure a total
excitation and distances between different states, correspondingly,
\begin{equation}
E\equiv E_{n},~~~e\equiv E_{n}-E_{m},  \label{eee}
\end{equation}%
In this variables, the basic equation (\ref{rnn}) for the occupancies $%
\overline{\rho }(E,t)$ of adiabatic states is transformed as
\begin{equation}
\Omega(E)\frac{\partial \bar{\rho}(E,t)}{\partial t}=\sum_{i,j=1}^{N}\dot{q}%
_{i}(t)\int_{0}^{t}dt^{\prime }K_{ij}(t,t^{\prime })\dot{q}_{j}(t^{\prime })%
\frac{\partial }{\partial E}\left[ \Omega(E)\frac{\partial \bar{\rho}%
(E,t^{\prime })}{\partial E}\right] .  \label{rnonm}
\end{equation}%
In Eq.~(\ref{rnonm}), the memory kernel, $K_{ij}(t,t^{\prime })$, is defined
as
\begin{equation}
K_{ij}(t,t^{\prime })=\frac{\sigma _{0,ij}^{2}(E,q+q^{\prime })}{\Gamma _{ij}%
}Y_{ij}(|q-q^{\prime }|)\int_{\infty }^{+\infty }de\ f_{ij}(|e|/\Gamma_{ij}
) R[|e|\Omega(E)]\mathrm{cos}(e[t-t^{\prime }]/\hbar ).  \label{Kts}
\end{equation}%
The integration limits over the energy spacing $e$ in Eq. (\ref{Kts}) were
extended to infinities since the time changes of the occupancy $\bar{\rho}%
(E,t)$ of the given state with the energy $E$ are mainly due to the direct
interlevel transitions from the close--lying states located at the distances
$|e|\ll E$. The explicit form of the two--level correlation function $R(x)$
in Eq. (\ref{Kts}) depends on the statistical ensemble of levels \cite%
{pame83}. It can be established the following result for three often used
ensembles:

(i) Gaussian Orthogonal Ensemble (\textrm{GOE})
\begin{equation}
R_{\mathrm{GOE}}(x)=1-\left( \frac{\mathrm{sin}(\pi x)}{\pi x}\right)
^{2}+\left( \int_{0}^{1}dy\frac{\mathrm{sin}(\pi xy)}{y}-\frac{\pi }{2}%
\right) \left( \frac{\mathrm{cos}(\pi x)}{\pi x}-\frac{\mathrm{sin}(\pi x)}{%
(\pi x)^{2}}\right) ,  \label{RGOE}
\end{equation}

(ii) Gaussian Unitary Ensemble (\textrm{GUE})
\begin{equation}
R_{\mathrm{GUE}}(x)=1-\left( \frac{\mathrm{sin}(\pi x)}{\pi x}\right) ^{2},
\label{RGUE}
\end{equation}

(iii) Gaussian Symplectic Ensemble (\textrm{GSE})
\begin{equation}
R_{\mathrm{GSE}}(x)=1-\left( \frac{\mathrm{sin}(2\pi x)}{2\pi x}\right)
^{2}+\int_{0}^{1}dy\frac{\mathrm{sin}(2\pi xy)}{y}\left( \frac{\mathrm{cos}%
(2\pi x)}{2\pi x}-\frac{\mathrm{sin}(2\pi x}{(2\pi x)^{2}}\right) ,
\label{RGSE}
\end{equation}%
where $x\equiv|E_{n }-E_{m}|\Omega(E_{n})$.

The main difference between the statistics is its behavior of $R(x)$ at
small energy spacings $x$. For the \textrm{GOE} statistics one has the
linear repulsion between levels, $R_{\mathrm{GOE}}\sim x$, the \textrm{GUE}
statistics implies the quadratic level repulsion, $R_{\mathrm{GUE}}\sim x^{2}
$, while in the \textrm{GSE} case we have $R_{\mathrm{GSE}}\sim x^{4}$. On
the other hand, $R_{\mathrm{GOE}}$, $R_{\mathrm{GUE}}$\ and $R_{\mathrm{GSE}}
$ are similar at moderate spacings $x$, when the spectral correlations
between levels consistently disappear, see also Refs.~\cite{goel65,koab08}.

Note that the dynamic process (\ref{rnonm}) may be treated as a quantum
mechanical diffusion of energy in space of the occupancies of
quantum adiabatic states, where a function
\begin{equation}
W(E,t)\equiv \Omega (E)\bar{\rho}(E,t),~~~~~\int_{E_{0}}^{+\infty
}W(E,t)dE=1,  \label{FEt}
\end{equation}%
gives a probability density to find the intrinsic quantum system with an
excitation energy lying in the interval $[E,E+dE]$\ at the moment of time $t$.
The time features of the non--Markovian quantum diffusive dynamics (\ref%
{rnonm})--(\ref{Kts}) are defined by the relationship between characteristic
time scales, $\tau_{ij} \sim \hbar /\Gamma_{ij} $ (caused by the finite
width $\Gamma_{ij}$ of the coupling matrix elements' energy distribution $%
f_{ij}(E)$) and the typical time interval $\tau _{\mathrm{macr},ij}$ of the
macroscopic variables' variations. For nuclear many--body Hamiltonians,
showing statistical properties, the energy distribution of the Breit--Wigner
shape \cite{zelev95}, $f_{ij}(|e|/\Gamma_{ij} )=(1/\pi )/(1+[|e|/\Gamma_{ij} ]^{2})$.
This gives rise to an exponentially decaying with time memory kernel (\ref%
{Kts}) (here we restrict ourselves by the one-dimension case to simplify the
notation)
\begin{equation}
K(t,t^{\prime })=\sigma _{0}^{2}(E,q+q^{\prime })\mathrm{exp}\left( -\frac{%
|t-t^{\prime }|}{\tau }\right)  \label{Kts2}
\end{equation}%
with $\tau =\hbar /\Gamma $. If one now assumes that $\tau $\ is the
shortest time scale in a system, one easily obtains the Markovian limit of
the quantum diffusion process (\ref{rnonm}) in terms of the probability
density function of the intrinsic quantum system (\ref{FEt}). Namely,
\begin{equation}
\frac{\partial W(E,t)}{\partial t}=-\frac{\partial }{\partial E}\left[
r(E,t)W(E,t)\right] +\frac{\partial ^{2}}{\partial E^{2}}\left[ D(E,t)W(E,t)%
\right] ,~~~~~\frac{\hbar }{\Gamma }\ll \tau _{\mathrm{macr}},  \label{Fet}
\end{equation}%
where the drift coefficient $r(E,t)$\ is equal to
\begin{equation}
r(E,t)=\frac{dD(E,t)}{dE}+\frac{D(E,t)}{\Omega (E)}\frac{d\Omega (E)}{dE}
\label{rEt}
\end{equation}%
and the diffusion coefficient $D(E,t)$ is given by
\begin{equation}
D(E,t)=(\hbar /\Gamma )\cdot \sigma _{0}^{2}(E,q)\dot{q}^{2}.  \label{DEt}
\end{equation}%
The drift term in the diffusion equation (\ref{Fet}) leads to
parametric excitation of the intrinsic quantum system that becomes possible
either due to the energy--dependence of the distribution of the slopes $%
\sigma _{0}^{2}(E)$\ in Eq.~(\ref{DEt}), or due to the average
level--density $\Omega (E)$ that grows with an intrinsic excitation $E$. The
diffusion coefficient $D(E,t)$ (\ref{DEt}) depends quadratically on the
velocity $\dot{q}$ of the parametric driving as a result of the assumed
weak--coupling regime \cite{kora10}. It should be pointed out that the
quantum--mechanical diffusion in space of many--body states' occupancies
gets stronger dependence on the velocity of the driving in the case of
Landau--Zener transitions between states (see Subsection \ref{LZ}).

\subsection{ Energy rate}
\label{rate}

To obtain equations of motion for the macroscopic collective parameters $%
q(t) $, we first find the average energy of the many--body system, $\mathcal{%
E}(t)=\mathrm{Tr}[\hat{H}\{q(t)\}\hat{\rho}(t)]$. Calculating the time
change of $\mathcal{E}(t)$, we obtain
\begin{equation}
\frac{d\mathcal{E}(t)}{dt}=\sum_i \dot{q}_i\frac{\partial E_{\mathrm{0}}(q)}{%
\partial q_i}+\sum_i\dot{q}_i\sum_{nm}\left( \frac{\partial \hat{H}}{%
\partial q_i}\right) _{mn}\rho _{nm}+\sum_{n}E_{n}\frac{\partial \rho _{nn}}{%
\partial t}+ \sum_i\dot{q}_i \sum_{n}\left( \frac{\partial \hat{H}}{\partial
q_i}\right) _{nn}\rho _{nn}.\quad  \label{dEdt}
\end{equation}

The first term in the right--hand side of Eq.~(\ref{dEdt}) describes a
change of the macroscopic potential energy $E_{\mathrm{0}}(q)=E_{\mathrm{pot}%
}(q)$. The second contribution to the energy rate $d\mathcal{E}/dt$ is
defined by the non--diagonal components of the density matrix $\rho _{nm}(t)$.
Its time evolution is caused by the virtual transitions among the
adiabatic states. Such a term is a microscopic source for the appearance of
the macroscopic kinetic energy. To demonstrate that, we write it as
\begin{eqnarray}  \label{Evt1}
\left( \frac{d\mathcal{E}}{dt}\right) ^{\mathrm{virt}}\equiv \sum_i \dot{q}%
_i\sum_{nm}\left( \frac{\partial \hat{H}}{\partial q_i}\right) _{mn}\rho
_{nm}  \notag \\
=\sum_i \dot{q}_i(t) \xi_i(t)+ 2\sum_{i} \dot{q}_i
\sum_j\sum_{nm}\int_{0}^{t}dt^{\prime } \mathcal{V}_{ij,nm}(t,t^{\prime })%
\dot{q}_j(t^{\prime }) [\rho _{mm}(t^{\prime })-\rho _{nn}(t^{\prime })],
\notag \\
\end{eqnarray}%
where
\begin{equation}
\xi_i(t)=\left( \frac{\partial \hat{H}}{\partial q_i}\right) _{mn}\rho
_{nm}(t=0),~~~ \mathcal{V}_{ij,nm}(t,t^{\prime
})=h_{i,nm}(t)h_{j,mn}(t^{\prime })\frac{\mathrm{cos}(\omega
_{nm}[t-t^{\prime }])}{\omega _{nm}}.  \label{V}
\end{equation}%
We formally extend the lower limit of the time integration in Eq.~(\ref{Evt1}%
) to $-\infty $. In this way, we would like to study stationary dynamics of
the complex quantum system, i. e., when the dynamics of the system does not
depend on the choice of initial time. It should be pointed out that this
does not imply a loss of any possible memory effects in collective motion.
Thus, integrating by parts the time integral in the r.h.s of Eq.~(\ref{Evt1}%
), one can show that
\begin{equation}
\int_{-\infty }^{t}dt^{\prime }\mathcal{V}_{ij,nm}(t,t^{\prime })\dot{q}%
_j(t^{\prime })[\rho _{mm}(t^{\prime })-\rho _{nn}(t^{\prime })]\approx
\sum_{l=0}^{+\infty }\omega _{nm}^{-(2l+3)}\times \frac{d^{(2l+1)}\left(
\dot{q}_j h_{i,nm}h_{j,mn}[\rho _{mm}-\rho _{nn}]\right) }{dt^{(2l+1)}}.
\label{tparts}
\end{equation}%
In the weak--coupling limit~\cite{kora10}, we obtain
\begin{equation}
\left( \frac{d\mathcal{E}}{dt}\right) ^{\mathrm{virt}}\approx \sum_i \dot{q}%
_i(t) \xi_i(t)+\sum_{i} \dot{q}_i \sum_j (B_{ij}(q)\ddot{q}_j+ \sum_k \frac{%
\partial B_{ij}(q)}{\partial q_k}\dot{q}_j \dot{q}_k),  \label{dEdtvir}
\end{equation}%
where the term
\begin{equation}
B_{ij}(q)=\sum_{n,m}h_{i,nm}h_{j,mn}\omega _{nm}^{-3}[\rho _{mm}-\rho _{nn}]
\label{B}
\end{equation}%
can be associated with a macroscopic inertia tensor.

The third term on the right--hand side of Eq.~(\ref{dEdt}) is determined by
the real transitions between the adiabatic states $\Psi _{n}(q)$ and it
defines how the energy of macroscopic motion is transferred into the energy
of the intrinsic excitations of the quantum system:
\begin{equation}
\left( \frac{d\mathcal{E}}{dt}\right) ^{\mathrm{real}}=\sum_i \dot{q}_i(t)
\sum_j \int_{0}^{t}dt^{\prime }K(t,t^{\prime })\dot{q}_j (t^{\prime
})\int_{E_0}^{+\infty } dE~E\frac{\partial }{\partial E}\left(\Omega(E)\frac{%
\partial \bar{\rho} (E,t^{\prime })}{\partial E}\right),  \label{dEdtreal}
\end{equation}%
where Eq.~(\ref{rnonm}) was used.

The fourth term in the r.h.s of Eq.~(\ref{dEdt}) is given by the
distribution of slopes of the adiabatic eigenstates $E_{n}$. Within the
random matrix model, the negative and positive slopes of the adiabatic
states are assumed to be equally distributed. Therefore, under the averaging
over all random realizations of the random matrices, modeling the nuclear
many body spectrum, one can neglect the contribution from the fourth term in
the r.h.s. of Eq.~(\ref{dEdt}).

Rewriting the time change of the average energy of the nuclear many--body
system (\ref{dEdt}) as
\begin{eqnarray}
\frac{d\mathcal{E}(t)}{dt}=\sum_i \dot{q}_i(t)F_i(q,t)  \notag \\
\equiv \sum_i \dot{q}_i(t) \big( \frac{\partial E_{\mathrm{pot}}}{\partial
q_i} +\xi_i(t)+\sum_j \left[ B_{ij}\ddot{q}_j+ \sum_k \frac{\partial B_{ij}}{%
\partial q_k}\dot{q}_j\dot{q}_k \right]  \notag \\
+\sum_j \int_{0}^{t}dt^{\prime }K_{ij}(t,t^{\prime })\dot{q}_j(t^{\prime })
\int_{E_{0}}^{+\infty }dE\Omega (E)E\frac{\partial }{\partial E} \left[
\Omega (E)\frac{\partial \bar{\rho}(E,t^{\prime })}{\partial E}\right] \big)
\label{Fi}
\end{eqnarray}
and assuming that all the partial contributions $F_i(q,t)$ to the energy
rate (having a meaning of the forces acting on the macroscopic collective
subsystem) are equal zero, we obtain transport description of the
macroscopic collective dynamics
\begin{eqnarray}  \label{eqm}
\sum_j \left[ B_{ij}\ddot{q}_j+ \sum_{k} \frac{\partial B_{ij}}{\partial q_k}%
\dot{q}_j\dot{q}_k \right]= -\frac{\partial E_{\mathrm{pot}}}{\partial q_i}
\notag \\
-\sum_j \int_{0}^{t}dt^{\prime }K_{ij}(t,t^{\prime })\dot{q}_j(t^{\prime })
\int_{E_{0}}^{+\infty }dE E\frac{\partial }{\partial E}\left[
\Omega (E) \frac{\partial \bar{\rho}(E,t^{\prime })}{\partial E}\right]%
-\xi_i(t).  \notag \\
\end{eqnarray}%

\subsection{Fluctuation--dissipation theorem}

The transport equations~(\ref{eqm}) for the classical collective
parameters $q(t)$ should be considered selfconsistently
with the equation of motion for the occupancies $\bar{\rho}(E,t)$
of the quantum many--body states (\ref{rnonm}). As was stated above
(see Eq.~(\ref{Fet}) and comment to it), the intrinsic quantum system
is excited by time variations of the collective parameters $q(t)$
provided that the averaged density of the many--body states $\Omega(E)$
grows with the excitation $E$. Importantly that such a parametric
excitation is time--irreversible in the sense that the intrinsic
quantum system (\ref{rnonm}) does not go back to the initial state
with time--reversing of the macroscopic collective parameters but
continues to be excited. This is so because under the parametric
time variations (defined by the absolute value of the parameters'
velocity $|\dot{q}(t)|$, see Eqs.~(\ref{rEt}) and (\ref{DEt})),
quantum--mechanical transitions to higher--lying states occur
more often than to lower--lying states. In turn, due to the
energy conservation condition (\ref{dEdt}), the increase
of the intrinsic excitation
\begin{equation}
E^*(t)\equiv \int_{E_{0}}^{+\infty }dE \Omega (E)E\bar{\rho}(E,t)
\label{E*}
\end{equation}
of the nuclear many--body system may be intrepreted as a corresponding
decrease (dissipation) of an energy associated with the time variations
of the macroscopic collective parameters $q(t)$:
\begin{equation}
E_{\rm coll}\equiv \sum_{i,j}\frac{1}{2}B_{i,j}(q)\dot{q}_i\dot{q}_j+E_{\rm pot}(q).
\label{Ecoll}
\end{equation}
In this way, we justified microscopically dissipative character of the nuclear
collective motion. For quite high initial excitations of the nucleus $E^*(t=0)$,
when the averaged density of many--body states is given by
\begin{equation}
\Omega(E^*(t=0)) \sim e^{E^*(t=0)/T},
\label{OmegaE}
\end{equation}
the collective dissipation is determined by memory kernels
\begin{equation}
\mathcal{K}_{ij}(t,t')=\frac{\sigma _{0,ij}^{2}(E^*(t=0),q[t]+q[t'])}{T}
\mathrm{exp}\left(-\frac{|t-t'|} {\tau_{ij}}\right) ,  \label{gamma}
\end{equation}%
of the retarded friction forces in the transport equations of motion
\begin{equation}
\sum_j \left[ B_{ij}\ddot{q}_j+ \sum_{k} \frac{\partial B_{ij}}{\partial q_k}%
\dot{q}_j\dot{q}_k \right]= -\frac{\partial E_{\mathrm{pot}}}{\partial q_i}
-\sum_{j,k} \frac{\partial B_{ij}}{\partial q_k}\dot{q}_j\dot{q}_k -\sum_j
\int_{0}^{t}dt^{\prime }\mathcal{K}_{ij}(t,t^{\prime })\dot{q}_j(t^{\prime
})-\xi_i(t),  \label{eqm1}
\end{equation}
In Eqs.~(\ref{OmegaE})--(\ref{eqm1}), a quantity $T$ is considered as
thermodynamic temperature of the nucleus.

To complete the transport description of the macroscopic collective
dynamics (\ref{eqm1}) one has to define statistical properties of the
terms $\xi_i(t)$ (\ref{V}), which are proportional to the non--diagonal
components $\rho_{nm}$ of the density matrix operator $\hat{\rho}(t)$
at initial moment of time $t=0$. If the initial distribution of the
diagonal components $\rho_{nn}$ is uniquely set in by the initial
excitation energy of the nuclear many--body system $E^*(t=0)$
(\ref{E*}), the initial values of the non--diagonal components
$\rho_{nm}$ are left without specifying. For each fixed value
of the initial excitation $E^*(t=0)$ , one has a buch of
the values $\rho_{nm}(t=0)$ which may be treated as random
numbers. In this case, we can treat the terms $\xi_i(t)$
in the transport
equations (\ref{eqm1}) as random forces acting on the
macroscopic collective parameters $q_i(t),~i=1,2,...$.
By using Eqs.~(\ref{V}),(\ref{hh}) and (\ref{gamma}),
we obtain for correlation functions of the random forces
$\xi_i(t)$,
\begin{equation}
\langle \xi_i (t)\xi_j (t^{\prime })\rangle
=T\mathcal{K}_{ij}(t,t^{\prime }),  \label{xixi}
\end{equation}
where it was assumed for simplicity that $|\rho_{nm}(t=0)|^2=1$.
The result (\ref{xixi}) represents the classical
fluctuation--dissipation relation between
dissipative and fluctuating properties of the macroscopic
collective dynamics both caused by the quantum--mechanical
diffusion of energy in the space of the occupancies of
many--body states (\ref{rnonm}).

\section{Application to large--amplitude collective motion}

\subsection{Descent from the fission barrier}

We are going to apply the developed transport approach (\ref{eqm1})--(\ref%
{xixi}) to the nuclear fission in two dimension space of shape variables $%
\{q_{1},q_{2}\}$. We consider the symmetric fission of heavy nuclei
whose shape is obtained by rotation of a profile function $\mathcal{L}(z)$
around $z$--axis. The Lorentz parameterization for $\mathcal{L}(z)$ is used
in the following form \cite{hamy88},
\begin{equation}
\mathcal{L}^{2}(z)=(z^{2}-\zeta _{0}^{2})(z^{2}+\zeta _{2}^{2})/Q\,,
\label{shape}
\end{equation}%
where the multiplier $Q$ guarantees the volume conservation,
\begin{equation}
Q=-[\zeta _{0}^{3}({\frac{1}{5}}\zeta _{0}^{2}+\zeta _{2}^{2})]/R_{0}^{3}\,.
\end{equation}%
Here, all quantities of the length dimension are expressed in the $R_{0}$
units, where $R_{0}$ is the radius of equal-sized sphere. The shape
variables $q_{1}=\zeta _{0}$ and $q_{2}=\zeta _{2}$ are related to the
nuclear elongation, $\zeta _{0}$, and the neck radius $\rho _{\mathrm{neck}%
}=\zeta _{2}/\sqrt{\zeta _{0}(\zeta _{0}^{2}/5+\zeta _{2}^{2})}.$ The basic
equations of motion (\ref{eqm1}) in two dimensions is written as%
\begin{equation}
\sum_{j=1}^{2}\left[ B_{ij}(q)\ddot{q}_{j}+\sum_{k=1}^{2}\frac{\partial
B_{ij}(q)}{\partial q_{k}}\ \overset{\cdot }{q_{j}}\overset{\cdot }{q_{k}}%
\right] =-\frac{\partial E_{\mathrm{pot}}(q)}{\partial q_{i}}%
-\sum_{j=1}^{2}\int_{0}^{t}dt^{\prime }\mathcal{K}_{ij}(t,t^{\prime })\dot{q}%
_{j}(t^{\prime })-\xi _{i}(t).  \label{eqmo2}
\end{equation}%
The numerical calculations are performed for the symmetric fission of the
nucleus $^{236}\mathrm{U}$ at temperature $T=2\ \mathrm{MeV}$. Considering
the nuclear descent from the saddle point to the scission, we use the
potential energy of deformation $E_{\mathrm{pot}}(\zeta _{0})$ from Ref.
\cite{hamy88} assuming that $\partial E_{\mathrm{pot}}(q_{1},q_{2})/\partial
\rho _{\mathrm{neck}}=0$. The scission line is derived from the condition of
the instability of the nuclear shape with respect to any variations of the
neck radius:
\begin{equation}
\left. \frac{\partial ^{2}E_{\mathrm{pot}}(q_{1},q_{2})}{\partial \rho _{%
\mathrm{neck}}^{2}}\right\vert _{\mathrm{scis}}=0  \label{inst}
\end{equation}%
Here, we wish to study the random force effect on the non--Markovian
dynamics (\ref{eqmo2}). To define properly a scission condition, we first
introduce a deterministic (i. e., without the random force) path of the
system (\ref{eqmo2}) and evaluate the nuclear neck radius at the scission
point $\left( \rho _{\mathrm{neck}}^{\mathrm{det}}\right) _{\mathrm{scis}}$.
Then, a bunch of $2\times 10^{4}$ stochastic trajectories $\zeta
_{0}(t),\zeta _{2}(t)$ is stopped as far as
\begin{equation}
\rho _{\mathrm{neck}}(\zeta _{0}(t),\zeta _{2}(t))=(\rho _{\mathrm{neck}}^{%
\mathrm{det}})_{\mathrm{scis}}.  \label{scis1}
\end{equation}%
As a result of that, we get a distribution of moments of time, $t_{\mathrm{sc%
}}$, when the system reaches the scission point (\ref{inst}). The
corresponding probability density $p$ of the scission events is shown in
figure 1 for the case of quite weak memory effects $\tau _{1}=2\times
10^{-23}~s$ and fairly strong memory effects $\tau _{2}=8\times 10^{-23}~s$.
For comparison, it is also shown (by vertical lines) the corresponding
scission times in the absence of the random force.

\begin{figure}[h]
\includegraphics[width=7cm]{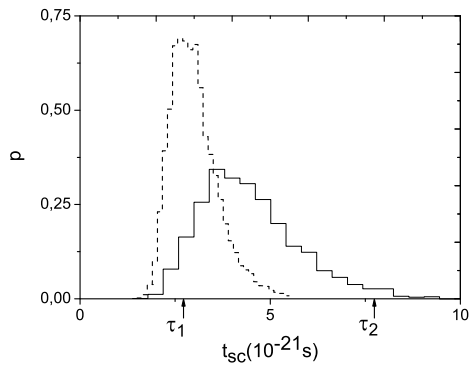}
\centering
\caption{Histogram, showing a probability density $p$ of moments of time $t_{%
\mathrm{sc}}$, when the stochastic trajectories $\protect\zeta _{0}(t),%
\protect\zeta _{2}(t)$ (\protect\ref{eqmo2}) hit the scission line $\protect%
\rho _{\mathrm{neck}}(\protect\zeta _{0},\protect\zeta _{2})=\left( \protect%
\rho _{\mathrm{neck}}^{\mathrm{det}}\right) _{\mathrm{scis}}$, is given at
two values of the memory time $\protect\tau $. The dashed histogram is found
for\textbf{\ $\protect\tau _{1}=2\times 10^{-23}~\mathrm{s}$ }(when the
memory effects in the system are quite weak) and the solid histogram
corresponds to the memory time $\protect\tau _{2}=8\times 10^{-23}~s$ (when
the memory effects are fairly strong). The corresponding times of descent in
the absence of the random force are given by small vertical arrows. Taken
from Ref. \protect\cite{kora09}.}
\end{figure}

We see that with the growth of the memory effects in (\ref{eqmo2}) the
distribution of the scission times $t_{\mathrm{sc}}$ becomes wider and the
centroid of the distribution shifts to the left compared to the
deterministic values of the scission time. One can interpret this as a
"stochastic" acceleration of the nuclear descent from fission barrier,
caused by the presence of the random force term in equations of motion (\ref%
{eqmo2}),\ see also Ref. \cite{kora09}.

Note that within the two--dimensional non--Markovian Langevin approach
(\ref{eqmo2})--(\ref{inst}) one can also calculate other characteristics
of nuclear fission process. Thus, in Ref.~ \cite{kora09} we have calculated
two experimentally observable quantities like the mean value and variance of
the total kinetic energy of fission fragments at infinity. By that, we have
estimated the value of memory time and found that $\tau\approx 8\times 10^{-23}~s$.

\subsection{Escape rate problem within the Langevin approach}

To measure the role of memory effects in collective dynamics of the nuclear
system on the way from ground state to saddle point, we restrict ourselves
by considering a one--dimensional collective motion $q(t)$ over a schematic
parabolic barrier shown in figure 2. The potential energy $E_{\mathrm{pot}}$
presents a single--well barrier formed by a smoothing joining at $q=q^{\ast
} $ of the potential minimum oscillator with the inverted oscillator
\cite{hofm97}
\begin{equation}
E_{\mathrm{pot}}=
\begin{cases}
\frac{1}{2}B\omega _{A}^{2}(q-q_{A})^{2},~~~~q\leq q^{\ast},
\\
=E_{\mathrm{pot,B}}-\frac{1}{2}B\omega _{B}^{2}(q-q_{B})^{2},~~~~q>q^{\ast}.
\end{cases}
\label{Epot}
\end{equation}

\begin{figure}[h]
\includegraphics[width=7cm]{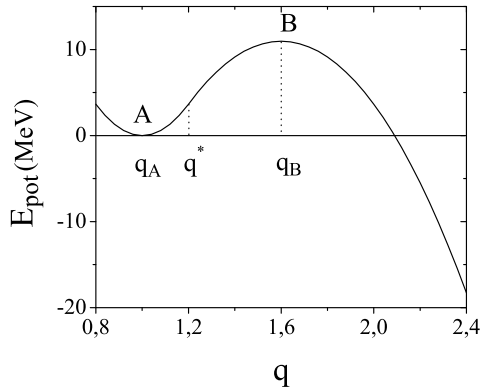}
\centering
\caption{Dependence of the potential energy $E_{\mathrm{pot}}$ (Kramers'
potential) on the shape parameter $q$.}
\end{figure}

We also adopt constant value $B$ for the collective inertia tensor (\ref{B})
and for the energy distribution $\sigma^2_0$ of the coupling matrix elements
in (\ref{gamma}), leading us to the equation of motion for the collective
variable $q(t)$,
\begin{equation}
B\ddot{q}=-\frac{\partial E_{\mathrm{pot}}}{\partial q}- \mathcal{K}%
_0\int_{0}^{t}dt^{\prime }\mathrm{exp}\left(-\frac{|t-t^{\prime }|}{\tau}%
\right) \dot{q}(t^{\prime })-\xi (t),  \label{eqm1_1}
\end{equation}%
where $\mathcal{K}_0=\sigma^2_0/T$ and
\begin{equation}
\langle \xi (t)\xi (t^{\prime })\rangle =T\mathcal{K}_{0}\mathrm{exp}\left( -%
\frac{|t-t^{\prime }|}{\tau }\right),  \label{xixi1}
\end{equation}%
see Eq.~(\ref{xixi}).

We studied the non--Markovian Langevin dynamics (\ref{eqm1_1}),
(\ref{xixi1}) by calculating a distribution of times $t_{\rm pre-sad}$
of the first crossing of the barrier top (pre--saddle times).
For that, the equation of motion (\ref{eqm1_1}) is
solved numerically by generating a bunch of the trajectories, all starting
at the potential well (point $A$ in figure 2) and having the
initial velocities distributed according to the Maxwell--Boltzman
distribution. On left panel of figure 3, it is shown the mean pre--saddle time
$\langle t_{\rm pre-sad} \rangle$ as a function of the memory
time $\tau$ and right panel of figure 3 gives the ration between
the standard deviation $\sigma(t_{\rm pre-sad})$ and the mean value
$\langle t_{\rm pre-sad} \rangle$ of the pre--saddle time distribution.

\begin{figure}[h]
\includegraphics[height=10cm,width=15cm]{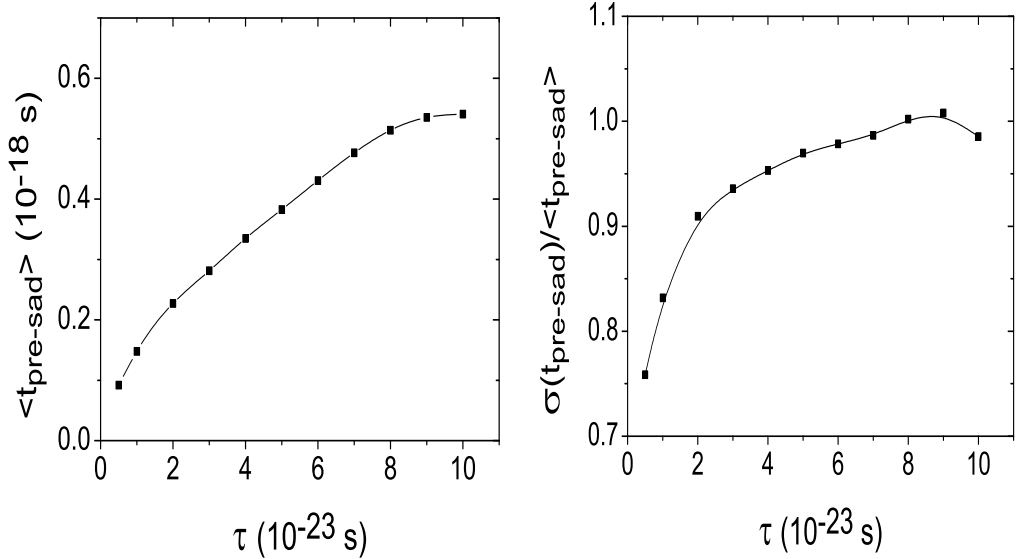}
\centering
\caption{The mean pre--saddle time $\langle t_{\mathrm{pre-sad}}\rangle$
(left panel) and the standard deviation $\sigma(t_{\rm pre-sad})$ over
$\langle t_{\rm pre-sad} \rangle$ (right panel) of the non--Markovian
diffusion dynamics (\protect\ref{eqm1_1}), (\protect\ref{xixi1}) are
shown as a function of the memory time $\protect\tau $.}
\end{figure}

An increase of the mean time $\langle t_{\mathrm{pre-sad}}\rangle$
of motion from the potential minimum $q_A$ to the saddle point $q_B$
with the memory time $\tau $
means that the memory effects in the Langevin dynamics (\ref{eqm1_1}),(\ref%
{xixi1}) hinders the diffusion over barrier. With growth of the memory time $%
\tau$ the adiabatic conservative force $-(\partial E_{\mathrm{pot}}/
\partial q)$ in equation of motion (\ref{eqm1_1}) gets an additional
contribution from the time--retarded force
\begin{equation}
-\mathcal{K} _{0}\int_{0}^{t}\mathrm{exp}\left( -\frac{|t-t^{\prime }|}{\tau
}\right) \dot{q}(t^{\prime })dt^{\prime }\rightarrow -\mathcal{K} _{0}\left[
q(t)-q_{A}\right] ,~~~~~\omega_B\tau \rightarrow \infty,  \label{tauinf}
\end{equation}
leading to the subsequent slowing down of motion to the saddle point $q_{%
\mathrm{b}}$. In the opposite limit of quite small values of the memory time
$\tau$, the slowing down of motion is exclusively due to effect from an
usual friction force since
\begin{equation}
-\mathcal{K}_{0}\int_{0}^{t}\mathrm{exp}\left( -\frac{|t-t^{\prime }|}{\tau }%
\right) \dot{q}(t^{\prime })dt^{\prime }\rightarrow -\mathcal{K} _{0}\tau
\dot{q}(t),~~~~~\omega_B\tau \rightarrow 0.  \label{tau0}
\end{equation}

Also note that with the growth of the memory time $\tau$, the
pre--saddle time distribution becomes wider as it is followed from
right panel of figure 3.

\subsection{Non--Markovian extension of the Kramers model}

One can also study the non--Markovian thermal diffusion over the parabolic
barrier (\ref{eqm1_1}), (\ref{xixi1}) with the help of an escape rate
characteristics, $R_{0}$, defining an exponential time decay of the
probability $\mathrm{Prob}(q[t] < q_{B})$,
\begin{equation}
\mathrm{Prob}(q[t] < q_{B})=e^{-R_{0}t},~~~~~t>1/\gamma _{0},  \label{prob}
\end{equation}
that the system $q[t]$ does not still reach the barrier top at $q_{B}$. Note
that the exponential decay with time of the survival probability (\ref{prob}%
) is reached at sufficiently large times $t$, when $t$ is larger than the
inverse characteristic friction coefficient $\gamma _{0}$, that will be
derived below.

To get an analytical estimate for the escape rate $R_{0}$, one can naturally
try to use the formalism of the Kramers' theory \cite{kram40} and that must
be extended to the non--Markovian system (\ref{eqm1_1}),(\ref{xixi1}). With
this purpose, we write down the equation of motion (\ref{eqm1_1}) for the
collective variable $q(t)$ in the vicinity of the barrier top $q_{B}$,%
\textbf{\ }
\begin{equation}
B\ddot{q}=B\omega _{B}^{2}(q-q_{B})-\mathcal{K}_{0}\int_{0}^{t}dt^{\prime }%
\mathrm{exp}\left( -\frac{|t-t^{\prime }|}{\tau }\right) \dot{q}(t^{\prime
})-\xi (t).  \label{eqm0}
\end{equation}%
General solution to the integro--differential equation (\ref{eqm0}) is
written as
\begin{equation}
q(t)=q_{B}+\mathcal{A}(t)q_{0}+\mathcal{B}(t)v_{0}-\int_{0}^{t}\mathcal{B}%
(t-t^{\prime })\xi (t^{\prime })  \label{coor}
\end{equation}%
or, in terms of velocity
\begin{equation}
v(t)=\dot{\mathcal{A}}(t)q_{0}+\dot{\mathcal{B}}(t)v_{0}-\int_{0}^{t}\dot{%
\mathcal{B}}(t-t^{\prime })\xi (t^{\prime })dt^{\prime },  \label{vel}
\end{equation}%
where $q_{0}$ is an initial coordinate and $v_{0}$ is an initial velocity of
the system. Two functions $\mathcal{A}(t)$ and $\mathcal{B}(t)$ are given by
\begin{equation}
\mathcal{A}(t)=1+\omega _{B}^{2}\int_{0}^{t}\mathcal{B}(t^{\prime
})dt^{\prime },~~~\mathcal{B}%
(t)=D_{1}e^{s_{1}t}+D_{2}e^{s_{2}t}+D_{3}e^{s_{3}t},  \label{AB}
\end{equation}%
where $s_{1},s_{2}$ and $s_{3}$ are roots of the secular equation
\begin{equation}
s^{3}+\frac{1}{\tau }s^{2}+\frac{\mathcal{K}_{0}-\omega _{B}^{2}}{B}s-\frac{%
\omega _{B}^{2}}{B\tau }=0  \label{seceq}
\end{equation}%
and the coefficients $D_{1},D_{2}$ and $D_{3}$ equal to
\begin{equation}
D_{1}=\frac{(s_{1}+1/\tau )}{(s_{1}-s_{2})(s_{1}-s_{3})},~~D_{2}=\frac{%
-(s_{2}+1/\tau )}{(s_{1}-s_{2})(s_{2}-s_{3})},~~D_{3}=\frac{(s_{3}+1/\tau )}{%
(s_{1}-s_{3})(s_{2}-s_{3})}.  \label{DDD}
\end{equation}%
The solutions to the secular equation (\ref{seceq}) have a threshold
behavior. For small enough memory times $\tau <\tau _{\mathrm{thresh}}$
(weak memory effects in the equation of motion (\ref{eqm0}), the first root $%
s_{1}$ of the secular equation (\ref{seceq}) is a positive number, while the
other two roots $s_{2}$ and $s_{3}$ are negative ones. In this case, the
mean collective path $\langle q(t)\rangle $ becomes to grow exponentially
with time as
\begin{equation}
\langle q(t)\rangle =a_{1}\mathrm{exp}(s_{1}t)+a_{2}\mathrm{exp}%
(-|s_{2}|t)+a_{3}\mathrm{exp}(-|s_{3}|t).  \label{qweak}
\end{equation}%
In the case of fairly large memory times $\tau >\tau _{\mathrm{thresh}}$
(strong enough memory effects), $s_{1}$ is still a positive number, whereas $%
s_{2}$ and $s_{3}$ become complex conjugated numbers, that results in
appearing of characteristic time oscillations of $\langle q(t)\rangle $:
\begin{equation}
\langle q(t)\rangle =a_{1}\mathrm{exp}(s_{1}t)+a_{4}\mathrm{exp}(-\mathrm{|Im%
}[s_{2}]|t)\mathrm{sin}(\mathrm{Re}[s_{2}]t)+a_{5}\mathrm{exp}(-|\mathrm{Im}%
[s_{2}]|t)\mathrm{cos}(\mathrm{Re}[s_{2}]t).  \label{qweak1}
\end{equation}

Since the random force $\xi (t)$ is Gaussian (but non--Markovian because of
Eq. (\ref{xixi1})), the two--dimensional process $\{q(t),v(t)\}$ (\ref{coor}%
)--(\ref{vel}) is also Gaussian and is defined by the probability
distribution function \cite{risk89},
\begin{equation}
\mathcal{W}(q\equiv q-q_{B},q_{0};v,v_{0};t)=\frac{1}{2\pi \sigma
_{q}(t)\sigma _{v}(t)\sqrt{1-r_{qv}(t)}}
\end{equation}%
\begin{equation}
\times \mathrm{exp}\left( -\frac{1}{2(1-r_{qv}^{2}(t))}\left\{ \frac{%
[q-\langle q(t)\rangle ]^{2}}{\sigma _{q}^{2}(t)}+\frac{[v-\langle
v(t)\rangle ]^{2}}{\sigma _{v}^{2}(t)}-\frac{2r_{qv}(t)[q-\langle
q(t)\rangle ][v-\langle v(t)\rangle ]}{\sigma _{q}(t)\sigma _{v}(t)}\right\}
\right) ,  \label{P}
\end{equation}%
where
\begin{equation}
\sigma _{q}(t)=\sqrt{\langle q^{2}(t)\rangle -\langle q(t)\rangle ^{2}}%
,~~~\sigma _{v}(t)=\sqrt{\langle v^{2}(t)\rangle -\langle v(t)\rangle ^{2}}%
,~~~r_{qv}(t)=\langle q(t)v(t)\rangle .  \label{ssr}
\end{equation}

All the time--dependent functions in (\ref{P}) can be expressed in terms of
the functions $\mathcal{A}(t)$ and $\mathcal{B}(t)$ (\ref{AB}) through the
relations (\ref{coor})--(\ref{AB}) and (\ref{xixi1}). Knowing the
probability distribution function $\mathcal{W}(q,q_{0};v,v_{0};t)$, one can
obtain the corresponding Fokker--Planck equation. We omit several
intermediate steps in deriving of the Fokker--Planck equation (for details,
see e.g. Ref. \cite{adel76}) and give the final result
\begin{eqnarray}
\left[ \frac{\partial }{\partial t}+v\frac{\partial }{\partial q}+\tilde{%
\omega} _{B}^{2}(t)q\frac{\partial }{\partial v}\right] \mathcal{W}(q \equiv
q-q_{B},q_{0};v,v_{0};t)=\frac{\gamma(t)}{B}\frac{\partial }{\partial v}[v%
\mathcal{W}]  \notag \\
+\frac{T\gamma(t)}{B^{2}}\frac{\partial ^{2}\mathcal{W}}{\partial v^{2}} +%
\frac{T}{B\omega _{B}^{2}}[\tilde{\omega}_B^{2}(t)-\omega _{B}^{2}]\frac{%
\partial ^{2}\mathcal{W}}{\partial v\partial q}.  \label{FPE}
\end{eqnarray}%
Here, the time--dependent friction coefficient $\gamma (t)$ and the
renormalized frequency parameter $\tilde{\omega}_{B}(t)$ of the parabolic
potential barrier are given by
\begin{equation}
\gamma (t)=B\frac{\ddot{\mathcal{A}}(t)\mathcal{B}(t)-\mathcal{A}(t)\ddot{%
\mathcal{B}}(t)}{\mathcal{A}(t)\dot{\mathcal{B}}(t)-\dot{\mathcal{A}}(t)%
\mathcal{B}(t)},~~~~~\tilde{\omega}_B^{2}(t)=\frac{\dot{\mathcal{A}}(t)\ddot{%
\mathcal{B}}(t)-\ddot{\mathcal{A}}(t)\dot{\mathcal{B}}(t)}{\mathcal{A}(t)%
\dot{\mathcal{B}}(t)-\dot{\mathcal{A}}(t)\mathcal{B}(t)}.
\label{tildegammaomega}
\end{equation}%
The non--Markovian character of the system shows up in the time--dependence
of the friction coefficient $\gamma $ and frequency parameter $\tilde{\omega}
_{B}^{2}$, and in the presence of a cross--term $\sim \partial ^{2}\mathcal{W%
}/\partial v\partial q$. The Markovian limit of the Fokker--Planck equation (%
\ref{FPE}) is reached at $\omega _{B}\tau \rightarrow 0$, when the retarded
force in Eq.~(\ref{eqm0}) simply turns to an ordinary friction force (\ref%
{tau0}). In general case, of course, the retarded force contains both the
time--dependent friction and conservative contributions,
\begin{equation}
-\mathcal{K}_{0}\int_{0}^{t}\mathrm{exp}\left( -\frac{|t-t^{\prime }|}{\tau }%
\right) \dot{q}(t^{\prime })\ dt^{\prime }=-\gamma (t)\dot{q}(t)+ B(\tilde{%
\omega}_{B}^{2}(t)-\omega _{B}^{2})(q(t)-q_{0}).  \label{split}
\end{equation}
We are looking for the stationary solution to Eq.~(\ref{FPE}) in the form
\begin{equation}
\mathcal{W}_{\mathrm{stat}}(q;v)=\mathrm{const}\cdot \mathfrak{F}(q,v)\cdot
\mathrm{exp}\left( -\frac{Bv^{2}/2}{T}-\frac{[E_{\mathrm{pot,B}}-B\tilde{%
\omega}_{B,\mathrm{sat}}^{2}q^{2}/2]}{T(1+\epsilon )}\right) ,
\label{statsol}
\end{equation}%
where $\epsilon =[\tilde{\omega}_{B,\mathrm{sat}}^{2}-\omega
_{B}^{2}]/\omega _{B}^{2}$, Substituting the solution (\ref{statsol}) into
Eq.~(\ref{FPE}), we obtain an equation for the function $\mathfrak{F}(q,v)$
\begin{equation}
(1+\epsilon )v\frac{\partial \mathfrak{F}}{\partial q}+\left( \frac{1}{%
1+\epsilon }\tilde{\omega}_{B,\mathrm{sat}}^{2}q+\frac{\gamma_0 v}{B}\right)
\frac{\partial \mathfrak{F}}{\partial v}=\frac{T\gamma_0}{B^{2}}\frac{%
\partial ^{2}\mathfrak{F}}{\partial v^{2}}+\frac{T\epsilon }{B}\frac{%
\partial ^{2}\mathfrak{F}}{\partial v\partial q}.  \label{Feq}
\end{equation}%
In Eqs.~(\ref{statsol}) and (\ref{Feq}), $\gamma_0$ and $\tilde{\omega}_{B,%
\mathrm{sat}}^{2}$ are long time values of the corresponding quantities (\ref%
{tildegammaomega}) taken at $\omega _{B}t\gg 1$. We found that
\begin{equation}
\gamma_0=-2B(s_{1}+\mathrm{min}(s_{2},s_{3})),~~~~~ \tilde{\omega}_{B,%
\mathrm{sat}}^{2}=-(s_{1}-\mathrm{min}(s_{2},s_{3})), ~~~~~~~\tau<\tau _{%
\mathrm{thresh}}.  \label{gammaomegatau0}
\end{equation}%
To be precise, at $\tau >\tau _{\mathrm{thresh}}$ the quantities $\gamma_0$
and $\tilde{\omega}_{B,\mathrm{sat}}^{2}$ do not exist as far as the
friction coefficient $\gamma(t)$ and frequency parameter $\tilde{\omega}%
_{B}^{2}(t)$ are strongly oscillating functions of time, taking positive as
well as negative values. The oscillations of $\gamma(t)$ and $\tilde{\omega}%
_{B}^{2}(t)$ are caused by the structure of functions $\mathcal{A}(t)$ and $~%
\mathcal{B}(t)$, see Eqs. (\ref{tildegammaomega}) and (\ref{AB}). Formally,
we will use the quantities $\gamma_0$ and $\tilde{\omega}_{B,\mathrm{sat}%
}^{2}$ even in the case of $\tau >\tau _{\mathrm{thresh}}$ in view of the
fact that one actually needs not the values $\gamma_0$ and $\tilde{\omega}%
_{B,\mathrm{sat}}^{2}$ itself but some combination of them, that can be
properly defined at the long time $\omega _{B}t\gg 1$ limit.

Looking now for the solution of Eq.~(\ref{Feq}) in the form
\begin{equation}
\mathfrak{F}\equiv \mathfrak{F}(v-aq)\equiv \mathfrak{F}(\zeta )
\label{Fzeta}
\end{equation}%
and repeating all derivation's steps of the Kramers' theory, we get
\begin{equation}
\mathfrak{F}(\zeta )=\mathfrak{F}_{0}\int_{-\infty }^{\zeta }\mathrm{exp}%
\left[ -\frac{\{(1+\epsilon )a-\gamma_0/B\}}{[2T/B] \{\gamma_0/B-\epsilon a\}%
}\zeta ^{\prime 2}\right] d\zeta ^{\prime },  \label{Fsol}
\end{equation}%
with
\begin{equation}
\mathfrak{F}_{0}=\sqrt{\frac{\{(1+\epsilon )a- \gamma_0/B\}}{[2\pi
T/B]\{\gamma_0/B-\epsilon a\}}}  \label{F0}
\end{equation}%
and
\begin{equation}
a=\frac{1}{1+\epsilon }\left[ \sqrt{\frac{\gamma_0^{2}}{4B^{2}}+\tilde{\omega%
}_{B,\mathrm{sat}}^{2}}+\frac{\gamma_0}{2B}\right] .  \label{a}
\end{equation}%
Therefore, the escape rate $R_0$ over parabolic barrier is found as
\begin{equation}
R_0=\frac{\omega _{A}}{2\pi \omega _{B}}\left[ \sqrt{\frac{\gamma_0^{2}} {%
4B^{2}}+\tilde{\omega}_{B,\mathrm{sat}}^{2}} -\frac{\gamma_0}{2B}\right]
e^{-E_{\mathrm{pot,B}}/T}=\frac{\omega _{A}}{2\pi \omega _{B}}s_{1}e^{-E_{%
\mathrm{pot,B}}/T}  \label{R_s}
\end{equation}%
that differs from the standard Kramers' result,
\begin{equation}
R_{\mathrm{Kr}}=\frac{\omega _{A}}{2\pi \omega _{B}}\left[ \sqrt{\frac{%
\gamma_0 ^{2}}{4B^{2}}+\omega _{B}^{2}}-\frac{\gamma_0 }{2B}\right] e^{-E_{%
\mathrm{pot,B}}/T},  \label{R_Kr}
\end{equation}%
by memory effects' modification $\tilde{\omega}_{B,\mathrm{sat}}$ of the
frequency parameter $\omega _{B}$ at the top of barrier. We would like to
stress that the transition in Eq.~(\ref{R_s}) from a nonlinear combination
of the friction coefficient $\gamma_0$ and frequency parameter $\tilde{\omega%
}_{B,\mathrm{sat}}$ to the largest positive root $s_{1}$ of the secular
equation (\ref{seceq}) is mathematically precise at any size $\tau $ of the
memory effects, see discussion after Eq.~(\ref{gammaomegatau0}).

In figure 4, we show by solid spline--fitting curve the escape rate
$R_{0}$ derived numerically from the Langevin collective dynamics
(\ref{eqm1_1}), (\ref{xixi1}). Analytical estimate of $R_{0}$, given
by the non--Markovian
extension of the Kramers' rate formula (\ref{R_s}) is shown in figure 8 by
dashed line. For comparison, we also plotted in figure by dotted line the
standard Kramers' expression for the escape rate (\ref{R_Kr}) with the $\tau
$--dependent friction coefficient
\begin{equation}
\gamma _{0}(\tau )=\frac{\mathcal{K}_{0}\tau }{1+[\mathcal{K}_{0}/B]\tau ^{2}%
},  \label{gammatau}
\end{equation}%
that correctly covers two limiting situations (\ref{tau0}) and (\ref{tauinf}%
) of the retarded friction force in Eq.~(\ref{eqm1_1}).

\begin{figure}[h]
\includegraphics[width=7cm]{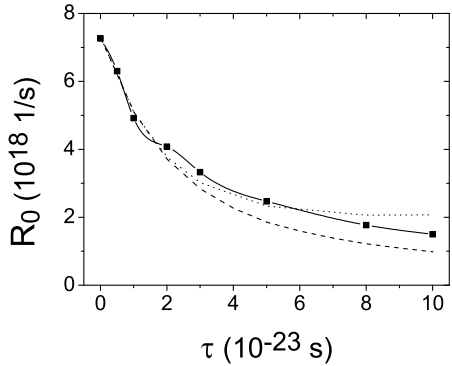}
\centering
\caption{Escape rate $R_0$ vs memory time $\protect\tau$. Solid
spline--fitting curve represents direct numerical determination of $R_0$ (%
\protect\ref{prob}) from the Langevin dynamics Eqs.~(\protect\ref{eqm1_1})
and (\protect\ref{xixi1}). Dashed line shows the result of the
non--Markovian extension of the Kramers' formula (\protect\ref{R_s}). The
improved formulation (\protect\ref{R_Kr})-- (\protect\ref{gammatau}) of the
standard Kramers' formula is given by dotted line.}
\end{figure}

One can see that the result (\ref{R_s}) correctly reproduces the decrease of
the rate $R$ of non--Markovian diffusion over barrier with the memory time $%
\tau $ (dashed line). On the other hand, the improved standard Kramers'
formula, given by Eqs.~(\ref{R_Kr}) and (\ref{gammatau}), overestimates the
value of $R_0$ at fairly large memory times $\tau$, see dotted line. From
that, one can conclude an importance of the taking into account the
memory--renormalization of the stiffness of the system near the top of a
potential barrier.

In the discussion above, we did not consider for a quantum tunneling
mechanism of the fission barrier penetration. This is justified by the fact
that the considered nucleus temperature of $2~{\rm MeV}$ is above the
characteristic crossover temperature $T_0$ \cite{grab84,hanggi85},
\begin{equation}
T_0=\frac{\hbar s_1}{2\pi},
\label{T0}
\end{equation}
which defines region where tunneling transitions dominate over thermally
activated transitions across the barrier. We found that $T_0\leq 0.13~{\rm MeV}$
at any value of the memory time $\tau$.

\subsection{Stochastic penetration over oscillating barrier}

One of the applications of the stochastic approach to the large amplitude
motion is the study of the features of the response of complex nonlinear
systems on periodic external field. The very famous example of such features
is the stochastic resonance phenomenon \cite{bsv81,mwr88}, when the response
of the nonlinear system on the harmonic perturbation is resonantly activated
under some optimal level of a noise. The resonant activation of the system
occurs when the frequency of the modulation is near the Kramers' escape rate
of the transitions from one potential well to another. The prototype of the
stochastic resonance studies is a model of overdamped motion between
potential wells of the bistable system. The frequency of the transitions
between wells is given by the Kramers' rate and the stochastic resonance is
achieved when a frequency of an external periodic modulation is of the order
of the Kramers' rate.

We start from the general Langevin formulation (\ref{eqm1_1}), (\ref{xixi1})
of the problem of diffusive overcoming of the potential barrier (\ref{Epot})
in the presence of a harmonic time perturbation:
\begin{equation}
B\ddot{q}(t)=-\frac{\partial E_{\mathrm{pot}}}{\partial q}-\mathcal{K}%
_{0}\int_{0}^{t}\mathrm{exp}\left( -\frac{|t-t^{\prime }|}{\tau }\right)
\dot{q}(t^{\prime })dt^{\prime }-\xi (t)+\alpha \cdot \mathrm{sin}(\omega t),
\label{lang2}
\end{equation}%
where $\alpha $ and $\omega $ are an amplitude and frequency of the
perturbation and the correlation function of the random force $\xi (t)$ is
given by Eq.~(\ref{xixi1}). Considering the diffusion over the barrier (\ref%
{lang2}),(\ref{xixi1}) in the presence of the external harmonic force, we
assume that the amplitude $\alpha $ of the periodic in time force $\alpha
\cdot \mathrm{sin}(\omega t)$ is so small that still the reaching the top of
the barrier is caused exclusively by diffusive nature of the process. In
figure 5, we show the typical dependencies of the mean pre--saddle time $%
t_{\mathrm{pre-sad}}$ on the frequency $\omega $ of the external harmonic
force. The calculations were performed for the weak, $\tau =2\times 10^{-23}~%
\mathrm{s}$, (lower curve in figure 5) and strong, $\tau =10\times 10^{-23}~%
\mathrm{s}$, (upper curve in figure 5) memory effects in the non--Markovian
diffusive motion over barrier.

\begin{figure}[h]
\includegraphics[width=7cm]{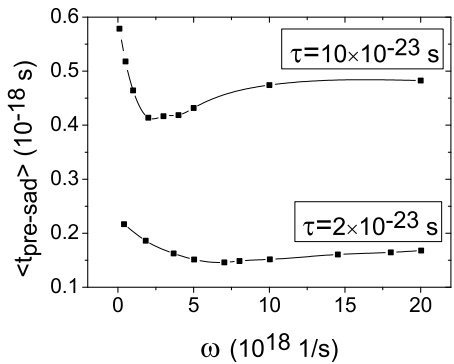}
\centering
\caption{The mean pre--saddle time $\protect \langle t_{\mathrm{pre-sad}}\rangle$ of the
non--Markovian diffusion process (\protect\ref{lang2}), (\protect\ref{xixi1}%
) is given as a function of the frequency $\protect\omega $ of the harmonic
time perturbation at two values of the correlation time $\protect\tau%
=2\times 10^{-23}~\mathrm{s}$ (lower curve) and $\protect\tau=10\times
10^{-23}~\mathrm{s}$ (upper curve).}
\end{figure}

In both cases the mean pre--saddle time $\langle t_{\mathrm{pre-sad}}\rangle$
non--monotonically depends on the frequency of the perturbation that is
character for the stochastic resonance phenomenon observed in a number of
different physical systems. From figure 5 one can conclude that the
diffusion over potential barrier in the presence of a harmonic time
perturbation is maximally accelerated at some definite so to say resonant
frequency $\omega _{\mathrm{res}}$ of the perturbation,
\begin{equation}
\omega _{\mathrm{res}}\approx \frac{1.5}
{\langle t_{\mathrm{pre-sad}}\rangle(\omega =0)}
\label{res1}
\end{equation}%
see also figure 2. In fact, the quantity $\langle t_{\mathrm{pre-sad}}\rangle(\omega =0) $
presents the characteristic time scale for the diffusion dynamics (\ref%
{lang2}). In the case of adiabatically slow time variations of the harmonic
force, $\omega \langle t_{\mathrm{pre-sad}} \rangle(\omega =0)\ll 1$ and
$t<\langle t_{\mathrm{pre-sad}}\rangle$,
one can approximately use $\alpha \cdot \mathrm{sin}(\omega t)\approx \alpha
\cdot \omega t$ and the diffusion over the barrier is slightly accelerated.
As a result of that, the mean pre--saddle time
$\langle t_{\mathrm{pre-sad}}\rangle (\omega)$ is smaller
than the corresponding unperturbed value
$\langle t_{\mathrm{pre-sad}}\rangle (\omega =0)$.
The same feature is also observed at the fairly large modulation's frequencies.
Thus, in the case of $\omega \langle t_{\mathrm{pre-sad}}\rangle(\omega =0)\gg 1$,
the harmonic perturbation $\alpha \cdot \mathrm{sin}(\omega t)$ may be treated
as a random noise term with the zero mean value and variance $\alpha ^{2}$.
Such a new stochastic term will lead to additional acceleration of the diffusion
over the barrier.

The existence of the resonant regime (\ref{res1}) in the
periodically modulated diffusion process (\ref{lang2}) is even more clear
visible through the time evolution of the survival probability $\mathrm{Prob}%
(q[t] < q_{B})$. For convience, in figure 6
we plotted $\mathrm{ln Prob}(q[t] < q_{B})$ for the resonant frequency
(\ref{res1}) $\omega =\omega _{\mathrm{res}}$ (curve 1), for fairly large
frequency $\omega=10\omega _{\mathrm{res}}$ (curve 2) and for quite
small frequency $\omega =\omega _{\mathrm{res}}/10$ (curve 3) of the
periodic time modulation.

\begin{figure}[h]
\includegraphics[width=7cm]{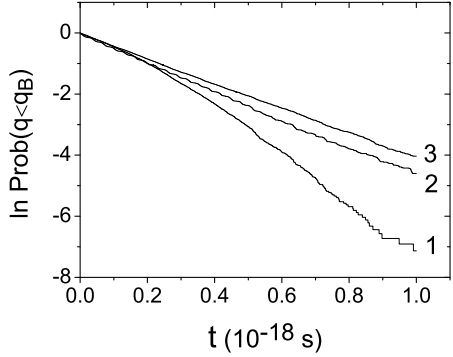}
\centering
\caption{Natural logarithm of the survival probability
$\protect \mathrm{ln Prob}(q[t] < q_{B})$ is given as a function
of time $t$ for the periodically modulated escape dynamics
(\protect\ref{lang2}), (\protect\ref{xixi1}).
Curve 1 corresponds to the modulation
on the resonance frequency (\protect\ref{res1})
$\protect\omega =\omega _{\mathrm{res}}$, curve 2 -- on the frequency
$\protect\omega =10\protect\omega _{\mathrm{res}}$ and curve 3 --
on the frequency $\protect\omega =\protect\omega _{\mathrm{res}}/10$.
The time dependencies are calculated for the memory time $\protect\tau%
=2\times 10^{-23}~\mathrm{s}$.}
\end{figure}

At very rare ($\omega =\omega _{\mathrm{res}}/10$) or frequent ($%
\omega =10\omega _{\mathrm{res}}$) periodic perturbations the probability
$\mathrm{Prob}(q[t] < q_B)$ of finding the system $q(t)$ (\ref{lang2}) on
the left from the saddle point $q_B$ exponentially decay with time and
characterized by almost constant escape (decay) rate $R_0$, see Eq.~(\ref{prob}).
On the other hand, periodic modulation of the non--Markovian diffusive
dynamics (\ref{lang2}), (\ref{xixi1}) at the resonance frequency
$\omega =\omega _{\mathrm{res}}$ provides much stronger decay with time
for the survival probability, $\mathrm{Prob}(q[t] < q_B)\sim {\rm exp}(-at^2)$.
In that case, the probability flow over the saddle point is
resonantly amplified.

\subsection{Diffusion of occupation probabilities within Landau-Zener
approach}

\label{LZ}

We restrict our analysis of the Landau-Zener approach to the one-dimensional
case and reduce the basic master equation (\ref{rnn}). To eliminate the
strong time oscillations of the observed quantitates $\rho _{nn}(t)$ etc. we
introduce the time averaging as
\begin{equation}
\bar{A}(t)={\frac{1}{\pi }}\int\limits_{-\infty }^{\infty }dt^{\prime }\,{%
\frac{\Delta }{{(t-t^{\prime 2}+\Delta ^{2}}}}\,A(t^{\prime }).
\end{equation}%
The master equation (\ref{rnn}) can be then presented in the following form
\cite{kolo95}

\begin{equation}
\frac{\partial \overline{\rho }_{nn}(t)}{dt}=\sum_{m}\overline{P}_{nm}(t)%
\left[ \overline{\rho }_{mm}(t)-\overline{\rho }_{nn}(t)\right] ,
\label{eq1}
\end{equation}%
where

\begin{equation}
\overline{P}_{nm}(t)=2\,\mathrm{Re}\int\limits_{o}^{\infty }dt^{\prime }\,%
\overline{D_{nm}(t)D_{mn}(t-t^{\prime })}\exp [i(\omega _{nm}+i\Gamma
_{nm})t^{\prime }]\quad  \label{p1}
\end{equation}%
and%
\begin{equation}
D_{nm}(t)=\left\langle \Psi _{n}\right\vert i\hbar {\frac{\partial }{{%
\partial t}}}\left\vert \Psi _{m}\right\rangle ,\quad \Gamma _{nm}=\Delta \,%
\dot{\omega}_{nm}.
\end{equation}

The transition rate (\ref{p1}) is a result of second order perturbation
theory in $\dot{q}$. However, the dimensionless cranking model parameter $%
\dot{q}|\left\langle \Psi _{n}\right\vert \partial \hat{H}/\partial
q\left\vert \Psi _{m}\right\rangle |/\hbar \omega _{mn}^{2}(q)$ is
increasing for two neighboring terms and therefore, the perturbation theory
criterion can not be fulfilled. We may improve the result (\ref{p1}) by
considering the Landau-Zener model \cite{zene32} of two crossing levels. In
the case of the Landau-Zener transitions the probability of a non-adiabatic
transition is
\begin{equation}
R^{\mathrm{LZ}}(e,\eta ).=\exp \left( -{\frac{{\pi e^{2}}}{{2\eta \mid \dot{q%
}\mid }}}\right) ,  \label{LZ1}
\end{equation}%
where $e$ is the gap size and $\eta $ is the slope of the avoided crossings
of two nearest levels. The energy difference $\Delta E$ of these levels as a
function of $q$ can be represented as
\begin{equation}
\Delta E=(e^{2}+\eta ^{2}q^{2})^{1/2}.
\end{equation}%
The highly excited energy levels $E_{n}$ in Eqs. (\ref{eq1}) and (\ref{p1})
exhibit many avoided crossings when the parameter $q$ is varied. We
introduce the average transition rate $\left\langle P^{\mathrm{LZ}%
}\right\rangle $ as the probability of transitions per unit time along of
the collective path $q(t)$ where the gap sizes $e$ and slopes $\eta $ are
randomly distributed
\begin{equation}
\left\langle P^{\mathrm{LZ}}\right\rangle =\dot{q}\int\limits_{0}^{\infty
}d\eta \int\limits_{0}^{\infty }de\ \>N(\eta ,e)\,R^{\mathrm{LZ}}(e,\eta ).
\label{LZ2}
\end{equation}%
Here $N(\eta ,e)\,d\eta \,de$ is the number of avoided crossings encountered
per unit length and slopes in the interval $[\eta ,\eta +d\eta ]$ and gap
sizes in the interval $[e,e+de]$. The distribution $N(\eta ,e)$ has been
calculated \cite{wilk89} in the limit where the gap $e$ is small (the
asymptotic slope $\eta $ is only well defined in this limit). Assuming that
the matrix elements $\left\langle \Psi _{n}\right\vert \partial \hat{H}%
/\partial q\left\vert \Psi _{m}\right\rangle $ are independently Gaussian
distributed, it was shown that
\begin{equation}
N(\eta ,e)=\Omega (E)\,\eta \,P(e)\,P^{\prime }(\eta ).  \label{LZ3}
\end{equation}%
The functions $P(e)$ and $P^{\prime }(\eta )$ are distribution functions of
gap sizes and slopes, respectively. In the case of chaotic systems the
distributions $P(e)$ and $P^{\prime }(\eta )$ depend on the spectral
statistics \cite{wilk89}.

Using the transition rate $\left\langle P^{\mathrm{LZ}}\right\rangle $ of (%
\ref{LZ2}) we can reduce the master equation (\ref{eq1}) into a diffusion
equation. Summation of both sides of Eq. (\ref{eq1}) over the states $n$
with $E_{n}<E$ gives,
\begin{equation}
\sum_{n\ (E_{n}<E)}{\frac{\partial f_{n}}{{\partial t}}}=-\sum_{n(E_{n}<E)}%
\left\langle P_{nm}^{\mathrm{LZ}}\right\rangle (f_{n}-f_{m}),  \label{f1}
\end{equation}%
where $f_{n}=\bar{\rho}_{nn}$ is the occupation probability. Let us
introduce $l(E)$ as the flux of probability from a level with energy less
than $E$ to levels with energy greater than $E$,
\begin{equation}
l(E)=\sum_{n(E_{n}<E),\ m(E_{m}>E)}\left\langle P_{nm}^{\mathrm{LZ}%
}\right\rangle (f_{n}-f_{m})
\end{equation}%
\begin{equation}
=\int\limits_{0}^{E}de_{1}\,\Omega (e_{1})\int\limits_{E}^{\infty
}de_{2}\,\Omega (e_{2})\left\langle P^{\mathrm{L.Z.}}(e_{1},e_{2})\right%
\rangle \ [f(e_{1})-f(e_{2})].  \label{lE1}
\end{equation}%
Using the cut-off properties of the transition rate (\ref{LZ1}) we transform
the flux $l(E)$ as
\begin{equation}
l(E)=-\Omega (E)\,{\frac{\partial f}{{\partial E}}}\,D(E)  \label{lE2}
\end{equation}%
where $D(E)$ is the diffusion coefficient
\begin{equation}
D(E)=\dot{q}\,\Omega (E)\int\limits_{0}^{\infty }de\
\>e^{2}\,P(e)\int\limits_{0}^{\infty }d\eta \>\ \eta \,P^{\prime }(\eta
)\,\exp \left( -{\frac{\pi e^{2}}{{2\eta \mid \dot{q}\mid }}}\right) .
\label{DE1}
\end{equation}%
Finally, we obtain the basic diffusion equation
\begin{equation}
\Omega (E)\,{\frac{\partial f}{{\partial t}}}={\frac{\partial }{{\partial E}}%
}\,\Omega (E)\,D(E)\,{\frac{\partial f}{{\partial E}}}.  \label{DE2}
\end{equation}%
This equation gives the Landau-Zener's evolution of the occupation
probabilities which is caused by slowly varying set of macroscopic variables
$q(t)$.

The distribution functions $P(e)$ and $P^{\prime }(\eta )$ can be evaluated
dependently on the spectral statistics:

(i) For systems with time-reversal invariance (Gaussian orthogonal ensemble)
\begin{equation}
P(e)\approx {\frac{\pi ^{2}}{{6}}}\Omega (E)\qquad (\mathrm{for}\>\ \mathrm{%
small}\text{ \ }\>e),  \label{GOE1}
\end{equation}%
\begin{equation}
P^{\prime }(\eta )={\frac{\eta }{{4\sigma ^{2}}}}\exp \{-\eta ^{2}/8\sigma
^{2}\}.  \label{GOE2}
\end{equation}

(ii) For systems without time-reversal invariance (Gaussian unitary
ensemble)
\begin{equation}
P(e)\approx {\frac{\pi ^{2}}{6}}\,e\,\Omega ^{2}(E)\qquad (\mathrm{for}\>\
\mathrm{small}\text{ \ }\>e),  \label{GUE1}
\end{equation}%
\begin{equation}
P^{\prime }(\eta )={\frac{\eta }{{2\sqrt{\pi }}\sigma ^{3}}}\,\exp \{-\eta
^{2}/4\sigma ^{2}\}.  \label{GUE2}
\end{equation}

We will calculate the rate of dissipation of energy due to Landau-Zener
transitions. Let us introduce the jump probability per unit time into an
energy interval $[e,e+de]$ as
\begin{equation}
\mathcal{P}(e)\,de=\dot{q}\int\limits_{0}^{\infty }d\eta \,N(\eta
,e;q)\,\exp \left[ -{\frac{{\pi e^{2}}}{{2\mid \dot{q}\mid \eta }}}\right]
\,de,  \label{Pe1}
\end{equation}%
where $N(\eta ,e;q)$ is defined in Eq. (\ref{LZ3}). The dissipation energy, $%
E_{\mathrm{diss}}$, is connected with the internal transitions which
accompany the macroscopic collective motion. The definition of $E_{\mathrm{%
diss}}$ in terms of the jump probability $P(e)$ depends on the total energy $%
E$ of the nucleus. We will consider two limiting cases.

(i) \underline{$E\approx E_{\mathrm{eq}}$} (motion close to the ground state)

The dissipation rate averaged over the spectral statistics is defined in
this case as
\begin{equation}
\left\langle \dot{E}_{\mathrm{diss}}\right\rangle
_{E}\,=\int\limits_{0}^{\infty }de\>\ e\>\mathcal{P}(e)  \label{diss1}
\end{equation}%
Using the expressions (\ref{Pe1}), (\ref{LZ3}) and (\ref{GOE1})-(\ref{GUE2})
we obtain
\begin{equation}
\mathrm{GOE}:\,\,\,\,\left\langle \dot{E}_{\mathrm{diss}}\right\rangle
_{E}\,\,=\mathrm{const}\cdot \Omega ^{2}(E)\,\sigma ^{2}(E)\,\dot{q}^{2},
\label{E1}
\end{equation}%
\begin{equation}
\mathrm{GUE}:\,\,\,\,\left\langle \dot{E}_{\mathrm{diss}}\right\rangle
_{E}\,\,=\mathrm{const}\cdot \Omega ^{3}(E)\,\sigma ^{5/2}(E)\,\mid \dot{q}%
\mid ^{5/2}.  \label{E2}
\end{equation}%
Thus, the friction force is proportional to $\dot{q}$ for the case of $%
\mathrm{GOE}$ and is proportional to $\mid \dot{q}\mid ^{3/2}$ for $\mathrm{%
GUE}$ statistics. This result is greatly reduced in the case of high
excitation energy.

(ii) \underline{$E\gg E_{\mathrm{eq}}$} (high excitation energy regime)

The definition of $\left\langle E_{\mathrm{diss}}\right\rangle _{E}$ is
different in this case from that of Eq. (\ref{diss1}) since transitions with
$e<0$ appear. We have,
\begin{equation}
\left\langle \dot{E}_{\mathrm{diss}}\right\rangle
_{E}\,=\int\limits_{-\infty }^{\infty }de\ e\>\mathcal{P}(e)  \label{diss2}
\end{equation}%
This expression and (\ref{Pe1}), (\ref{LZ3}) and (\ref{GOE1})-(\ref{GUE2})
lead to
\begin{equation}
\mathrm{GOE}:\,\,\,\,\left\langle \dot{E}_{\mathrm{diss}}\right\rangle
_{E}\,=\mathrm{const}\cdot \Omega (E)\,\frac{d\Omega (E)}{dE}\,\sigma
^{5/2}(E)\,\mid \dot{q}\mid ^{5/2},  \label{E3}
\end{equation}%
\begin{equation}
\mathrm{GUE}:\,\,\,\,\left\langle \dot{E}_{\mathrm{diss}}\right\rangle
_{E}\,=\mathrm{const}\cdot \Omega ^{3}(E)\,\frac{d\Omega (E)}{dE}\,\sigma
^{3}(E)\,\mid \dot{q}\mid ^{3}.  \label{E4}
\end{equation}

The main feature of this result is the dependence of the dissipation energy
on the derivative $d\Omega(E)/dE$ of the level density. This means that time
irreversible exchange between the collective and internal degrees of freedom
is possible if the phase space volume is increasing with the excitation
energy of the system.

\section{Non-Markovian Langevin--like dynamics of Fermi--liquid drop}

The large amplitude motion of a nucleus can be studied in terms of fluid
dynamic approaches which allow us to reduce the problems of the avoided
crossings of\ adiabatic energies\ $E_{n}(q)$, see e.g.
Eq. (\ref{B}) for the inertia tensor $B(q)$. In
general, the nuclear fluid dynamics is influenced strongly by the Fermi
motion of nucleons and is accompanied by the dynamic distortion of the
Fermi-surface in momentum space \cite{kosh04,kolo83}. The presence of the
dynamic Fermi-surface distortion gives rise to some important consequences
in the nuclear dynamics which are absent in classical liquids. The dynamics
of a nuclear Fermi liquid is determined by the pressure tensor instead of
the scalar pressure as in a classical liquid. This fact changes the
conditions for the propagation of the isoscalar and isovector sound
excitations and creates a strong transverse component in the velocity field
of the particle flow. Furthermore, because of the Fermi-surface distortion,
the scattering of particles on the distorted Fermi-surface becomes possible
and the relaxation of collective motion occurs \cite{abkh59}. The equations
of motion of nuclear Fermi-liquid take then a non-Markovian form. The memory
effects depend here on the relaxation time and provide a connection between
both limiting cases of the classical liquid (short relaxation time limit)
and the quantum Fermi-liquid (long relaxation time limit). The Markovian
dynamics only exist in these two limiting cases.

Under the description of the properties of a drop of quantum Fermi liquid,
one can start from the collisional kinetic equation for the distribution
function in phase space $f(\mathbf{r},\mathbf{p};t)$. Using the standard $%
\mathbf{p}$-moment procedure, one can derive the continuity equation for the
particle density $\rho $ and Euler--like equation for the velocity field $%
u_{\nu }$(for details, see Refs.~\cite{kora01,kosh04,kota81}),
\begin{equation}
{\frac{\partial }{\partial t}\rho }=-{\nabla }_{\nu }(\rho u_{\nu }),
\label{e1}
\end{equation}%
\begin{equation}
m\rho {\frac{\partial }{\partial t}}u_{\nu }+m\rho \ u_{\mu }\nabla _{\mu }\
u_{\nu }+\nabla _{\nu }\mathcal{P}+\rho \nabla _{\nu }{\frac{\delta \epsilon
_{\mathrm{pot}}}{\delta \rho }}=-\nabla _{\mu }P_{\nu \mu }^{\prime },
\label{e2}
\end{equation}%
where $\mathcal{P}$ is the isotropic part of the pressure, $P_{\nu \mu
}^{\prime }$ is its anisotropic part caused by the dynamic distortion of
Fermi surface and $\epsilon _{\mathrm{pot}}$ is the potential energy density
of inter-particle interaction. The pressure tensor $P_{\nu \mu }^{\prime }$
is derived by the distribution function in phase space $f(\mathbf{r},\mathbf{%
p};t)$. Considering the dynamic Fermi--surface distortions up to
multipolarity $l=2$ one can establish \cite{kosh04,kolo83} the following
equation the pressure tensor $P_{\nu \mu }^{\prime }$
\begin{equation}
{\frac{\partial }{\partial t}}P_{\nu \mu }^{\prime }+\mathcal{P\ }\left(
\nabla _{\nu }u_{\mu }+\nabla _{\mu }u_{\nu }-{\frac{2}{3}}\delta _{\nu \mu
}\nabla _{\alpha }u_{\alpha }\right) =I_{\nu \mu }+Y_{\nu \mu },  \label{2.5}
\end{equation}%
where $I_{\nu \mu }$ is the second moment of the collision integral $%
I[\delta f]=-\delta f/\tau $:
\begin{equation}
I_{\nu \mu }={\frac{1}{m}}\int {\frac{d\mathbf{p}}{(2\pi \hbar )^{3}}}p_{\nu
}p_{\mu }I[\delta f]  \label{2.8}
\end{equation}%
and $Y_{\nu \mu }$ gives the contribution from the random force
\begin{equation}
Y_{\nu \mu }={\frac{1}{m}}\int {\frac{d\mathbf{p}}{(2\pi \hbar )^{3}}}p_{\nu
}p_{\mu }Y.  \label{2.9}
\end{equation}%
The equation of motion (\ref{e1})-(\ref{2.5}) are closed with respect to the
velocity field $u_{\nu }$.

For an incompressible and irrotational flow of the nuclear fluid with a
sharp surface, the local equation of motion (\ref{e1})-(\ref{2.5}) can be
reduced to the equations for the variables $q_i(t),~i=\overline{1,N}$
that specify the shape of the nucleus \cite{kora01}. The continuity equation
(\ref{e1}) has to be complemented by the boundary condition on the moving
nuclear surface $S$. Below we will assume that the axially symmetric shape
of the nucleus is defined by rotation of the profile function $\rho =%
\mathcal{L}(z,\{q_{i}(t)\})$ around the $z$-axis in the cylindrical
coordinates $\rho ,z,\varphi $. The velocity of the nuclear surface is then
given by \cite{ivko95}
\begin{equation}
u_{S}=\sum_{i=1}^{N}\bar{u}_{i}\dot{q}_{i},  \label{u1}
\end{equation}%
where
\begin{equation}
\bar{u}_{i}=({{\partial }\mathcal{L}/{\partial q_{i}}})/\Lambda ,\ \ \
\Lambda =\sqrt{1+({{\partial }\mathcal{L}/{\partial z}})^{2}}.  \label{u2}
\end{equation}%
The potential of the velocity field takes the form
\begin{equation}
\phi =\sum_{i=1}^{N}{}_{i}\overline{\phi }\ \dot{q}_{i},  \label{u3}
\end{equation}%
where the potential field $\overline{\phi }_{i}\equiv \overline{\phi }_{i}(%
\mathbf{r,}q)$ is determined by the equations of the following Neumann
problem
\begin{equation}
\nabla ^{2}\overset{\_}{\phi }_{i}=0\,,\,\,\ \ \ (\mathbf{n\cdot }\nabla
\overset{\_}{\phi }_{i})_{S}={\frac{1}{\Lambda }}{\frac{{\partial }\mathcal{L%
}}{{\partial q_{i}}}},  \label{aaa26}
\end{equation}%
where $\mathbf{n}$ is the unit vector which is normal to the nuclear
surface. Using Eqs. (\ref{e1}) and (\ref{e2}) with multiplying Eq. (\ref{e2}%
) by $\nabla _{\mu }\overline{\phi }_{i}$ and integrating over $\mathbf{r}$,
one obtains
\begin{equation}
\sum_{j=1}^{N}[B_{ij}(q)\ddot{q}_{j}+\sum_{k=1}^{N}\frac{\partial B_{ij}}{%
\partial q_{k}}\ \overset{\cdot }{q_{j}}\overset{\cdot }{q_{k}}%
+\int_{t_{0}}^{t}dt^{\prime }\exp (\frac{t^{\prime }-t}{\tau })\kappa
_{ij}(t,t^{\prime })\ \overset{\cdot }{q_{j}}(t^{\prime })]=-\frac{\partial
E_{\mathrm{pot}}(q)}{\partial q_{i}}.  \label{aaa27}
\end{equation}%
Here $B_{ij}(q)$ is the inertia tensor
\begin{equation}
B_{ij}(q)=m\rho _{0}\oint ds\bar{u}_{i}\ \overline{\phi }_{j}
\end{equation}%
and we ignore the random force. The memory kernel $\kappa _{i,j}(t,t^{\prime
})$\ in Eq. (\ref{aaa27}) is given by
\begin{equation}
\kappa _{ij}(t,t^{\prime })=2\ \int d\mathbf{r\ }\mathcal{P}(\mathbf{r,}%
q(t^{\prime }))\ \left[ \nabla _{\nu }\nabla _{\mu }\overset{\_}{\phi }_{i}(%
\mathbf{r,}q(t))\right] \left[ \nabla _{\nu }\nabla _{\mu }\overset{\_}{\phi
}_{j}(\mathbf{r,}q(t^{\prime }))\right] \ .  \label{kappa}
\end{equation}%
The potential field $\overline{\phi }_{i}\equiv \overline{\phi }_{i}(\mathbf{%
r,}q)$ are determined by a solution to the Neumann problem (\ref{aaa26}).

To solve Eq. (\ref{aaa27}) we will rewrite it as a set of two equations.
Namely,
\begin{equation}
\sum_{j=1}^{2}[B_{ij}(q)\ddot{q}_{j}+\sum_{k=1}^{2}\frac{\partial B_{ij}}{%
\partial q_{k}}\ \overset{\cdot }{q_{j}}\overset{\cdot }{q_{k}}]=-\frac{%
\partial E_{\mathrm{pot}}(q)}{\partial q_{i}}+R_{i}(t,q)  \label{main1}
\end{equation}%
and
\begin{equation}
\frac{\partial R_{i}(t,q)}{\partial t}=-\frac{R_{i}(t,q)}{\tau }%
+\sum_{j=1}^{2}\kappa _{ij}(q,q)\dot{q}_{j}\ \ \qquad \mathrm{at\qquad }%
R_{i}(t=0,q)=0,  \label{main2}
\end{equation}%
where $q=\{q_{1},q_{2}\}=\{\zeta _{0},\zeta _{2}\}$ and the terms $\sim \dot{%
q}_{i}\dot{q}_{j}$ were omitted in Eq. (\ref{main2}), as the next order
corrections. The memory kernel $\kappa _{ij}(q,q)$ is given by
\begin{equation}
\kappa _{ij}(q,q)=\frac{2}{5}m\rho _{0}v_{F}^{2}\ \int d\mathbf{r\ }(\nabla
_{\nu }\nabla _{\mu }\overset{\_}{\phi }_{i}(\mathbf{r,}q))\ (\nabla _{\nu
}\nabla _{\mu }\overset{\_}{\phi }_{j}(\mathbf{r,}q)).  \label{kappa2}
\end{equation}

We will apply this approach to the case of symmetric nuclear fission
described, assuming the Lorentz parameterization for the profile function $%
\mathcal{L}(z)$ in the form of Eq.~(\ref{shape}). We have performed
numerical calculation for symmetric fission of the nucleus $^{236}\mathrm{U}$.
We solved Eqs. (\ref{main1}) and (\ref{main2}) numerically using the
deformation energy $E_{\mathrm{pot}}(q)$ from Refs. \cite{hamy88}. The
scission line was derived from the condition of the instability of the
nuclear shape with respect to the variations of the neck radius, see Eq. (%
\ref{inst}). The equations of motion (\ref{main1}) and (\ref{main2}) were
solved with the initial conditions corresponding to the saddle point
deformation and the initial kinetic energy $E_{\mathrm{kin,0}}=1~\mathrm{MeV}
$ (initial neck velocity $\dot{\zeta _{2}}=0$).

\begin{figure}[htp]
\includegraphics[width=0.35\textwidth]{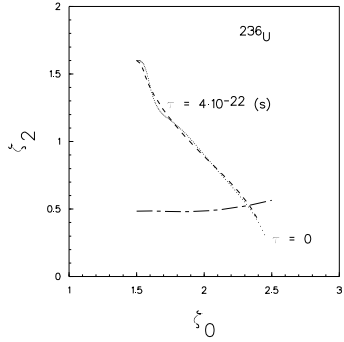}
\centering
\vspace{20mm}
\caption{Trajectories of descent from the saddle point of the nucleus
$^{236}$U in the $\protect \zeta _{0},\protect\zeta _{2}$ plane.
Dashed line represents the result of
the calculation in presence of the memory effects and dotted line is for the
case of Markovian (no memory) motion with the friction forces. We have used
the relaxation time $\protect\tau =4\times 10^{-22}~\mathrm{s}$ and the
initial kinetic energy $E_{\mathrm{kin}}=1\,\mathrm{MeV}$. Dot-dashed line
is the scission line derived from the condition (\protect\ref{inst}).}
\end{figure}

In Figure 7 we show the dependence of the fission trajectory, i.e.,
the dependence of the neck parameter $\zeta _{2}$ on the elongation $\zeta
_{0},$ for the fissioning nucleus $^{236}\mathrm{U}$ for two different
values of the relaxation time $\tau $: $\tau =4\times 10^{-22}~\mathrm{s}$
(dashed line) and $\tau =0\times 10^{-23}~{\rm s}$ (dotted line).
The scission line (dot-dashed
line in Figure  7) was obtained as a solution to Eq. (\ref{inst}). We
define the scission point as the intersection point of the fission
trajectory with the scission line. As can be seen from figure  11 the
memory effect hinders slightly the neck formation and leads to a more
elongated scission configuration. To illustrate the memory effect on the
observable values we have evaluated the translation kinetic energy of the
fission fragments at infinity, $E_{\mathrm{kin}}$, and the prescission
Coulomb interaction energy, $E_{\mathrm{Coul}}$. The value of $E_{\mathrm{kin%
}}$ is the sum of the Coulomb interaction energy at scission point, $E_{%
\mathrm{Coul}}$, and the prescission kinetic energy $E_{\mathrm{kin,ps}}$.
Namely,
\begin{equation}
E_{\mathrm{kin}}\mathrm{=}E_{\mathrm{Coul}}+E_{\mathrm{kin,ps}}\mathrm{.}
\label{hhh}
\end{equation}%

\begin{figure}[h]
\includegraphics[width=7cm]{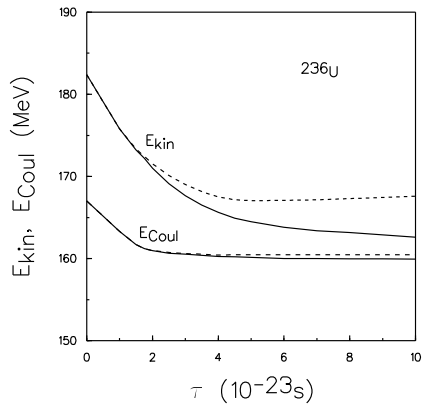}
\centering
\vspace{20mm}
\caption{Fission-fragment kinetic energy, $E_{\mathrm{kin}}$, (curves 1) and the
Coulomb repulsive energy at the scission point, $E_{\mathrm{Coul}}$, (curves
2) versus the relaxation time $\protect\tau $ for the nucleus $^{236}$U.
Solid lines represent the result of the calculation in presence of the
memory effects and dashed lines are for the case of Markovian (no memory)
motion with the friction forces. The initial kinetic energy is $E_{\mathrm{%
kin,0}}=1\,\mathrm{MeV}$.}
\end{figure}

The influence of the memory effects on the fission-fragment kinetic energy, $%
E_{\mathrm{kin}}$, and the prescission Coulomb interaction energy, $E_{%
\mathrm{Coul}}$, is shown in figure  8. As seen from figure 8,
the memory effects are neglected at the short relaxation time regime where
the memory integral is transformed into the usual friction force. In the
case of the Markovian motion with friction (dashed line), the yield of the
potential energy, $\Delta E_{\mathrm{pot}}$, at the scission point is
transformed into both the prescission kinetic energy, $E\mathrm{_{kin,ps},}$
and the time irreversible dissipation energy, $E_{\mathrm{dis}}$, providing $%
\Delta E_{\mathrm{pot}}=E\mathrm{_{kin,ps}}$ $+\ E_{\mathrm{dis}}\mathrm{.}$
In contrast to this case, the non-Markovian motion with the memory effects
(solid line) produces an additional time reversible prescission energy, $E_{%
\mathrm{F,ps}},$ caused by the distortion of the Fermi surface. In this
case, the energy balance reads $\Delta E_{\mathrm{pot}}=E\mathrm{_{kin,ps}}$
$+\ E_{\mathrm{dis}}+E_{\mathrm{F,ps}}.$ Note that the used parametrization
of the fissioning nucleus at the scission point, leads to the prescission
Coulomb energy $E_{\mathrm{Coul}}$ which is about 5 $\mathrm{MeV}$ lower
(for $^{236}\mathrm{U}$) than the Coulomb interaction energy of the scission
point shape \cite{dama77}. Taking into account this fact and using the
experimental value of the fission-fragment kinetic energy $E_{\mathrm{kin}}^{%
\mathrm{\exp }}=168$ $\mathrm{MeV}$ \cite{dama77}, one can see from
figure 12 that the Markovian motion with friction (dashed line) leads to the
overestimate of the fission-fragment kinetic energy $E_{\mathrm{kin}}.$ In
the case of the non-Markovian motion with the memory effects (solid line), a
good agreement with the experimental data is obtained at the relaxation time
of about $\tau =8\times 10^{-23}~\mathrm{s}$. A small deviation of the
prescission Coulomb energy $E_{\mathrm{Coul}}$ obtained at the non-Markovian
motion (solid line in figure 12) from the one at the Markovian motion
(dashed line in figure 12) is caused by the corresponding deviation
of both fission trajectories in Figure 8.

In figure 9 we illustrate the memory effect on the
saddle-to-scission time $t_{\mathrm{sc}}.$ In the case of the non-Markovian
motion (solid line), the delay in the descent of the nucleus from the
barrier grows with the relaxation time $\tau $ (at
$\tau \geq 4\cdot 10^{-23}~\mathrm{s}$). This is mainly due to
the hindering action of the elastic
force caused by the memory integral. The saddle-to-scission time increases
by a factor of about 2 due to the memory effect at the
{\it experimental} value of the relaxation time $\tau =8\times
10^{-23}~\mathrm{s}$ which was derived from the fit of the fission-fragment
kinetic energy $E_{\mathrm{kin}}$ to the experimental value of $E_{\mathrm{%
kin}}^{\mathrm{\exp }}$\ (see above).

\begin{figure}[h]
\includegraphics[width=7cm]{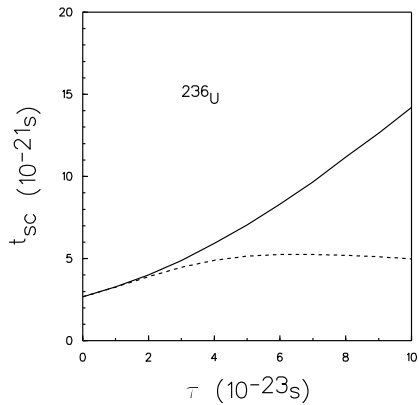}
\centering
\vspace{20mm}
\caption{ Dependence upon relaxation time $%
\protect\tau $ of the saddle-to-fission time, $t_{\mathrm{sc}}$, for the
descent from the barrier in the case of two-dimension ($\protect\zeta _{0},%
\protect\zeta _{2}$) parametrization for the nucleus $^{236}$U. Solid line
represents the result of the calculation in presence of the memory effects
and dashed line is for the case of Markovian (no memory) motion with the
friction forces. The initial kinetic energy is
$E_{\mathrm{kin}}=1\,\mathrm{MeV}$.}
\end{figure}

\section{Summary}

In attempt to understand microscopically the origin for dissipative and
fluctuating properties of collective modes of motion in nuclear many--body
systems, we have developed an approach, based on the Zwanzig's projection
technique (\ref{rnm})--(\ref{rnn}).

We have averaged the intrinsic nucleonic dynamics over suitably chosen
statistics of the randomly distributed matrix elements (\ref{hh}), measuring
the coupling between the intrinsic nucleonic and macroscopic collective $%
q(t) $ subsystems, and the energy spacings. In assumption of the weak
coupling \cite{kora10}, we have derived the diffusion--like equation of
motion (\ref{rnonm}) for the ensemble averaged occupancies of the adiabatic
many--body states, describing a process of non--Markovian energy diffusion
and that occurs due to quantum--mechanical transitions between the
many--body levels. Going beyond the weak--coupling limit \cite{kora10}, we
have also studied the energy diffusion process in space of the occupancies
of the adiabatic many--body states for Landau--Zener transitions between
levels (\ref{LZ1}), where the diffusive properties essentially depend on
quantum statistics of many--body states, see Eqs.~(\ref{E3}), (\ref{E4}). It
has been demonstrated that under the growing of the averaged density of
levels with energy $E$, the intrinsic nucleonic subsystem parametrically
excites, that can, in turn, be interpreted as corresponding dissipation of
collective energy due to constancy of the total energy of the nuclear
many--body system.

We have shown that memory effects in the diffusion equation (\ref{rnonm}) is
caused by finite spread $\Gamma $ of the randomly distributed coupling
matrix elements (\ref{hh}) and disappears for quite spread distributions,
when $\hbar /\Gamma $ is the shortest time scale in the system. We have
demonstrated that a (memory or correlation) time $\tau=\hbar/\Gamma $
defines non--Markovian character of macroscopic collective dynamics (\ref%
{eqm}) and also determines statistical properties of of the random force
term, incorporated into equations of motion for the collective variables (%
\ref{eqm}), through the fluctuation--dissipation theorem (\ref{xixi}).

We have further applied our non--Markovian Langevin approach (\ref{eqm}), (%
\ref{xixi}) to the description of descent of the nucleus from the fission
barrier. We have showed that the random force accelerate significantly the
process of descent from the barrier for both the Markovian and
non--Markovian Langevin dynamics, see figure 1. This fact may be explained
by the correlation properties (\ref{xixi}) of the random force in the
non--Markovian Langevin equations.

We have generalized the Kramers' theory of escape rate over parabolic
potential barrier (of figure 2) to systems with non--Markovian dynamics (\ref%
{eqm1_1}),(\ref{xixi1}). The found expression for the escape rate (\ref{R_s}%
) differs from the standard Kramers result (\ref{R_Kr}) by
memory--renormalization of the frequency parameter $\tilde{\omega}_{B,%
\mathrm{sat}}$ (\ref{tildegammaomega}) at the top of barrier. It has been
shown that $\tilde{\omega}_{B,\mathrm{sat}}$ is effectively smaller that the
frequency parameter $\omega _{B}$ of the adiabatic barrier, giving rise to
slowing down of the thermal diffusion over barrier with the increase of the
size $\tau$ of memory effects, see figure 3.

We have also studied the non--Markovian diffusion (\ref{lang2}), (\ref{xixi1}%
) over parabolic barrier in the presence of weak periodic time perturbation.
The mean pre--saddle time was calculated as a function of the frequency $%
\omega $ of the perturbation (see figure 5) and found that the diffusive
motion over the barrier may be significantly accelerated at some definite
(resonance) frequency $\omega _{\mathrm{res}}$, that is defined by the mean
time of motion in the absence of perturbation.


By use of $p$-moments techniques, we have reduced the collisional kinetic
equation to the equations of motion for the local values of particle
density, velocity field and pressure tensor, see Eqs.~(\ref{e1})--(\ref{2.5}).
To apply our approach to the nuclear large amplitude motion, we have
assumed that the nuclear liquid is incompressible and irrotational. We have
derived the velocity field (\ref{u1})--(\ref{aaa26}), which depends then on the
nuclear shape parameters $q(t)$ due to the boundary condition on the moving
nuclear surface. Finally, we have reduced the problem to a macroscopic
equation of motion for the shape parameters $q(t)$ (\ref{aaa27})--(\ref{kappa}).
Thus, we consider a change (not necessary small) of the nuclear shape
which is accompanied by a small quadrupole distortion of the Fermi
surface. The obtained equations of motion for the collective variables $q(t)$
contains the memory integral which is caused by the Fermi-surface distortion
and depends on the relaxation time $\tau $.

We have shown that the development of instability near the fission barrier
is strongly influenced by the memory effects if the relaxation time $\tau $\
is large enough. In this case, a drift of the nucleus from the barrier to
the scission point is accompanied by characteristic shape oscillations which
depend on the relaxation time $\tau $. The shape oscillations appear due to
the elastic force induced by the memory integral. The elastic force\ acts
against the adiabatic force $-\ \partial E_{\mathrm{pot}}(q)/\partial q$\
and hinders the motion to\ the scission point. In contrast to the case of
the Markovian motion, the delay in the fission is caused here by the
conservative elastic force and not only by the friction force. Due to this
fact, the nucleus loses a part of the prescission kinetic energy converting
it into the potential energy of the Fermi surface distortion instead of the
time-irreversible heating of the nucleus.

The memory effects lead to the decrease of the fission-fragment kinetic
energy, $E_{\mathrm{kin}},$ with respect to the one obtained from the
Markovian motion with friction, see Figure 7. This is because a
significant part of the potential energy at the scission point is collected
as the energy of the Fermi surface deformation. Note that the decrease of
the fission-fragment kinetic energy due to the memory effects is enhanced in
the\ rare collision regime (at larger relaxation time) while the effect due
to friction decreases. An additional source for the decrease of the
fission-fragment kinetic energy is caused by the shift of the scission
configuration to that with a larger elongation parameter $\zeta _{0},$ in
the case of the non-Markovian motion. Due to this fact, the repulsive
Coulomb energy of the fission fragments at the scission point in
figure 12 decreases with respect to the case of the Markovian motion.


\section{Acknowledgments}

This study was supported in part by the program Support
for the development of priority areas of scientific research
of the National Academy of Sciences of Ukraine,
Grant No. 0120U100434.


\end{document}